# Correlation between unconventional superconductivity and strange metallicity revealed by operando superfluid density measurements

Ruozhou Zhang[1,2]†, Mingyang Qin[1,3]†*, Chenyuan Li[4]†, Zhanyi Zhao[1,5]†, Zhongxu Wei[1,6], Juan Xu[1], Xingyu Jiang[1,5], Wenxin Cheng[1,5], Qiuyan Shi[1,5], Xuewei Wang[1,5], Jie Yuan[1,5,7], Yangmu Li[1,5], Qihong Chen[1,5,7], Tao Xiang[1,8], Subir Sachdev[4*], Zi-Xiang Li[1,5*], Kui Jin[1,5,7*], and Zhongxian Zhao[1,5,7]

[1]*Beijing National Laboratory for Condensed Matter Physics, Institute of Physics, Chinese Academy of Sciences, Beijing 100190, China*

[2]*State Key Laboratory of Surface Physics and Department of Physics, Fudan University, Shanghai 200433, China*

[3]*Department of Materials Science and Engineering, Southern University of Science and Technology, Shenzhen 518055, China*

[4]*Department of Physics, Harvard University, Cambridge MA 02138, USA*

[5]*School of Physical Sciences, University of Chinese Academy of Sciences, Beijing 100049, China*

[6]*Department of Physics, Southern University of Science and Technology, Shenzhen 518055, China*

[7]*Songshan Lake Materials Laboratory, Dongguan, Guangdong 523808, China*

[8]*Beijing Academy of Quantum Information Sciences, Beijing, 100193, China*

†These authors contributed equally to this work.

*: qinmy@sustech.edu.cn; sachdev@g.harvard.edu; zixiangli@iphy.ac.cn; kuijin@iphy.ac.cn

## Abstract

**Strange-metal behavior has been observed in superconductors ranging from cuprates to pressurized nickelates[1-7], but its relationship to unconventional superconductivity remains elusive. Here, we perform *operando* superfluid density measurements on ion-gated FeSe films. We observe for the first time a**

**synchronized evolution of superconducting condensate and the strange-metal phase with electron doping. A linear scaling between zero-temperature superfluid density and the strange-metal resistivity coefficient is further established, which nails down a direct link between the formation of superfluid in the superconducting state and the scattering of carriers in the strange-metal normal state. Remarkably, the scaling also applies for different iron-based and cuprate superconductors despite their distinct electronic structures and pairing symmetries. Such a correlation can be reproduced in a theoretical calculation on the two-dimensional Yukawa-Sachdev-Ye-Kitaev model[8] by considering a cooperative effect of quantum critical fluctuation and disorder. These findings indicate a fundamental principle governing superconducting condensation and strange-metal scattering in unconventional superconductors.**

**Main text**

In the search for the microscopic mechanism of unconventional superconductors, a concerted effort has been directed toward understanding their normal states and how they link to superconductivity. It has been found that the normal state of various unconventional superconductors, e.g., cuprates[1,2,9], iron-based materials[3,10], magic-angle graphene[7], and nickelates[4-6] exhibits a linear-in-temperature ($T$-linear) resistivity extending to the low temperature. Such an anomalous transport behavior[11], termed as the strange metal, is acknowledged to be intimately related to unconventional superconductivity[1,2,9,10,12]. However, despite extensive investigations over the past decades[8,13-16], the microscopic mechanism of strange-metal transport and the interplay between strange metal and superconductivity remain actively debated. One of the central questions is what the fundamental correlation between the strange-metal normal state and superconductivity is. To find clues for solving the puzzle, inspecting the crucial ingredient governing superconductivity across the strange-metal regime is immensely required.

One key parameter characterizing the superconducting state is the superfluid density,

$\rho_s$ ($\equiv \lambda_{ab}^{-2} \propto n_s/m^*$, with $\lambda_{ab}$ the in-plane magnetic penetration depth, $n_s$ the superconducting carrier density and $m^*$ the effective electron mass), which encodes the rigidity of quantum-mechanism phase of Cooper pairs, reflecting the resilience of pairing condensate against phase fluctuations[17,18]. It has been widely observed in cuprates that the superconducting transition temperature ($T_c$) scales with $\rho_{s0}$ [$\equiv \rho_s(T \rightarrow 0)$] (see e.g., Refs. [18-20]), which is beyond the standard Barden-Cooper-Schrieffer (BCS) scenario that $T_c$ is determined by the pairing strength. More intriguingly, in overdoped cuprates such as $Tl_2Ba_2CuO_{6+\delta}$ (Tl2201) and $La_{2-x}Sr_xCuO_4$ (LSCO), the normal-state carrier density increases with hole doping, but both $\rho_{s0}$ and the $T$-linear resistivity coefficient $A_1$ decrease[13,14], suggesting a pivotal role of the strange-metal state in the pairing condensate. Consequently, establishing a quantitative relationship between $\rho_{s0}$ and $A_1$, and exploring to what extent it can be applied to unconventional superconductors are of vital importance in understanding the essential physics underlying the connection between unconventional superconductivity and the strange metallicity, and establishing a unified theoretical framework of unconventional superconductors. However, the great challenges involve manipulating superconductivity minutely across the strange-metal regime and obtaining precise $\rho_{s0}$ data in an efficient way.

For LSCO, the systematic evolution of $\rho_{s0}$ with $T_c$ in the overdoped regime has been revealed by Božović *et al*[18], based on massive data from over 2000 films prepared within about 12 years. Other than chemical substitution methods, the ionic liquid gating (ILG) technique is an approach that can continuously tune $T_c$ of a single sample by virtue of either electrostatic or electrochemical effects[21]. FeSe, a prototype iron-based superconductor with the simplest crystal structure containing only the Fe-Se layer, has shown its unique compatibility with the ILG technique, i.e., the almost fivefold enhancement of $T_c$ under gating[22]. Importantly, the strange-metal state of FeSe films with $T$-linear resistivity, linear-in-field ($H$) magnetoresistance, and $H/T$ scaling of magnetoresistance was unambiguously identified, and its evolution with ionic-liquid gating was also disclosed in our previous work[10].

Here, we perform *operando* $\rho_s$ measurements by the two-coil mutual inductance (TCMI) technique on the ion-gated FeSe films, with $T_c$ finely tuned in a wide range of approximately 11 to 43 K in a single sample. With this high-efficiency approach, we uncover a systematic dependence of $T_c$ on $\rho_{s0}$ for the iron-based superconductors, i.e., $T_c \propto \rho_{s0}^{0.55 \pm 0.11}$. Intriguingly, we find that the superfluid condensate keeps in lockstep with the strange-metal phase during the gating process. A linear correlation between $\rho_{s0}$ and $A_1$, i.e., $\rho_{s0} \sim A_1$, is further quantified, which also works for iron pnictides and overdoped cuprates. Such a relation is well captured by the two-dimensional Yukawa-Sachdev-Ye-Kitaev (2d-YSYK) model with a spatially disordered fermion-scalar coupling and potential disorder. Our findings reveal the quantitative connection between unconventional superconductivity and the strange metallicity, which may be common to the physics of different unconventional superconductors. (In this text, the symbols " $\propto$ " and "~" denote a proportional relationship and a generic linear relationship allowing for a constant intercept, respectively.)

FeSe films ($T_c \approx 11$ K) with thicknesses of $60 \pm 5$ nm were deposited on (001)-oriented LiF substrates with size of $5 \times 5$ mm$^2$ by the pulsed laser deposition technique. The x-ray diffraction measurements confirm the high-quality (00*l*) oriented growth of the film (Supplementary Section 1). The pristine films were loaded in our ILG device as schematically shown in Fig. 1a. An ionic liquid *N*,*N*-diethyl-*N*-(2-methoxyethyl)-*N*-methyl ammonium bis(trifluoromethylsulfonyl)imide (DEME-TFSI) was used as the dielectric, covering both the FeSe film and the gate electrode. Under a positive gate voltage, the $H^+$ ions in the ionic liquid enter the FeSe film, which has been evidenced by the time-of-flight secondary ion mass spectroscopy measurement[10]. Such an electron doping effect brings forth a transition in the carrier type of FeSe from the coexistence of electrons and holes to high-density electrons only (Supplementary Section 2), which is accompanied by an enhancement of $T_c$ to approximately 40 K (Refs. [10,22]).

The superfluid density, $\rho_s \equiv \lambda_{ab}^{-2}$, is difficult to be accurately measured since $\lambda_{ab}$ is typically thousands of angstroms. The TCMI technique is a suitable method to obtain the absolute value of $\rho_s$ of superconducting films, which has attracted increasing attention due to its high sensitivity and flexibility[18]. In order to in situ monitor the evolution of $\rho_s$ of FeSe films, the ILG device was integrated into a homemade transmission-type TCMI apparatus that consists of a drive coil and a pickup coil[23]. Figure 1c shows the temperature-dependent pickup coil voltage, $V = V_x + iV_y$, for a gated state with $T_c \approx 43$ K. The real component, $V_x$, undergoes a rapid drop around $T_c$, originating from the diamagnetic effect of the screening supercurrent (see the calculated current distribution in the FeSe film in Fig. 1b). Correspondingly, the imaginary component, $V_y$, shows a sharp dip, resulting from the energy dissipation[24]. The absolute value of $\rho_s$ can be accurately extracted from $V$ by employing a self-developed fast wavelet collocation (FWC) method[25], as shown in Fig. 1d (see Supplementary Sections 3 and 4 about the device configuration and the $\lambda_{ab}$ measurements).

By minutely tuning the superconductivity of FeSe film via the ILG technique, a systematic evolution of the diamagnetic responses is obtained. It is shown that the diamagnetic signals exhibit a quasi-parallel shift to high temperature with gating (Supplementary Sections 5). Importantly, each curve exhibits only a single transition, confirming the bulk modulation of the superconductivity. Figure 2a shows the extracted temperature dependence of superfluid density, $\rho_s(T)$, for all gated states. A common feature is the flattening of $\rho_s$ at low temperatures, which becomes more prominent for higher-$T_c$ states. Such a feature has been observed in iron-based superconductors, e.g., $Ba_{1-x}K_xFe_2As_2$ (Ref. [26]), noted as an indication of the nodeless pairing gap. We fit the $\rho_s(T)$ data with nodeless gap equation, i.e., $\rho_s(T) = \rho_{s0}(1 - Ce^{-\Delta_{\min}/k_B T})$, where $\Delta_{\min}$ is the minimum gap size, $k_B$ is the Boltzmann constant, $C$ is the numerical parameter, and $\rho_{s0}$ is the superfluid density at zero-temperature limit, which works well at low temperatures (see Fig. 2b). We note that the $\rho_{s0}$ value of the pristine FeSe film ($1.09 \pm 0.09$ μm$^{-2}$) is consistent with the value of $FeSe_{1-x}Te_x$ film ($1.29 \pm 0.11$ μm$^{-2}$) measured by the scanning SQUID microscope[27], but is about 1/6 of the bulk FeSe (6.25 μm$^{-2}$, Ref. [28]). Such a relatively low ratio of thin film to bulk crystal seems ubiquitous in unconventional superconductors, e.g., 1/5 for $Ba(Fe_{1-x}Co_x)_2As$ (Refs. [29,30]) and 1/9 for $YBa_2Cu_3O_{6+y}$ (Refs. [20,24]), which was attributed to the relatively strong quantum phase fluctuations[31] and/or disorder effect[32] in thin films.

With the systematic $\rho_s$ data, we explore the dependence of $T_c$ on $\rho_{s0}$, which exhibits a simple power-law behavior (Fig. 2c), that is, $T_c = \gamma\rho_{s0}^{0.55 \pm 0.11}$, where $\gamma$ is a numeric parameter. Here, the uncertainty in the power-law exponent is mainly caused by the error bars of the $T_c$ value, i.e., $\Delta T_c$ (see Fig. 1c), which may be associated with the relatively inhomogeneous modulation of the superconductivity by ionic-liquid gating, or the strong superconducting fluctuations of FeSe[28]. Such a relationship between $T_c$ and $\rho_{s0}$ holds well against different gap symmetries used to extract $\rho_{s0}$ (Supplementary Section 6). We emphasize that this is the first systematic $T_c(\rho_{s0})$ scaling established in iron-based superconductors, benefiting from our high-efficiency *operando* measurements.

It was suggested that a power-law index ≈ 0.5 might occur in the dirty limit of a *d*-wave BCS superconductor[33,34]. However, such a simple 'dirty' BCS picture could not quantitatively explain our observations, since our data unambiguously violate the Homes' law, i.e., $\rho_{s0} \propto \sigma_{dc} T_c$ (where $\sigma_{dc}$ is the normal-state conductivity measured closed to $T_c$) (see Supplementary Section 7), that typically holds for a dirty BCS superconductors owing to the Glover–Ferrell–Tinkham sum rule[18,35,36]. We noticed that the $T_c \propto \rho_{s0}^{0.5}$ relation was also observed near the critical doping where superconductivity vanishes[18,37] and was attributed to the critical quantum phase fluctuations[20,38,39]. Besides, a similar scaling relation is found by our 2d-YSYK modeling of a quantum critical metal with fermion-scalar Yukawa couplings and potential disorder (Fig. S13b in Supplementary Section 14). Although further studies are needed to determine the mechanism responsible for the power-law behavior observed here, the strong correlation between $T_c$ and $\rho_{s0}$ demonstrates that phase rigidity plays a dominant role in determining $T_c$ of ion-gated FeSe, in contrast to the standard BCS paradigm where $T_c$ is bounded by the temperature at which Cooper pairs form. This is corroborated by large ratios of $(T_c/T_F, \Delta_{min}/E_F) \approx (0.11, 0.23)$ for samples with $T_c > 39$ K ($E_F = k_B T_F$ is the effective Fermi energy derived from $\rho_{s0}$), indicating that ion-gated FeSe locates in the BCS-Bose-Einstein condensation (BEC) crossover regime[28] (Fig. 2d).

More importantly, by displaying the systematic $\rho_s$ data in the phase diagram with varying electron doping and temperature, we are able to trace the evolution of the superconducting condensate with the normal-state resistivity (Fig. 3a). Intriguingly, we find that the magnitude of the superfluid density and the strength of the strange-metal scattering (dissipation) are synchronously enhanced with electron doping, which is accompanied by the similar increasing trends for $T_c$ (see the gray circles and gray squares) and the upper bound temperature of $T$-linear resistivity ($T_1$ defined in the inset of Fig. S7). These observations provide strong evidence that the formation of superfluid condensation is linked with the strange-metal normal state, which is compatible with the scenario recently proposed in overdoped cuprates that superfluid condensation originates from incoherent carriers contributing to the strange-metal transport[13,40]. To the best of our knowledge, such a visualized description of the interplay between unconventional superconductivity and the strange metallicity has not been established before.

Figure 3b shows both $\rho_{s0}$ and $A_1^{\square}$ as a function of electron doping for the ion-gated FeSe film, where $A_1^{\square}$ is the $T$-linear resistivity coefficient normalized by the distance between adjacent superconducting layers. Strikingly, $\rho_{s0}$ tracks $A_1^{\square}$ closely for all gated states, which unambiguously points to a linear correlation between $\rho_{s0}$ and $A_1^{\square}$, i.e., $\rho_{s0} = \alpha\widetilde{A}_1^{\square}$ with $\alpha = 0.43 \pm 0.02$ μm$^{-2}$ $\Omega^{-1}$K (Fig. 3c). Here, a constant has been subtracted from $A_1^{\square}$, namely $\widetilde{A}_1^{\square} = A_1^{\square} - A_0$, where $A_0 = 17.5$ Ω K$^{-1}$ represents the non-zero extrapolation of $A_1$ to $\rho_{s0} = 0$ μm$^{-2}$ or $T_c = 0$ K (Ref. [10]). Note that with the error bars taking into account, the linear relationship is the best fit to the $\rho_{s0}$ versus $A_1^{\square}$ data (Supplementary Section 9).

To figure out to what extent the linear scaling of ion-gated FeSe is applied to other superconducting materials, we collect the $\rho_{s0}$ and $A_1^{\square}$ data from previous reports (see the original data in Supplementary Section 11). Systematic $A_1^{\square}$ data of electron-doped cuprate $La_{2-x}Ce_xCuO_4$ (LCCO) have been obtained on composition-spread films[9], but corresponding $\rho_{s0}$ data are lacking due to the difficulty of accurately characterizing the local superfluid density. The calculated upper bound of $\rho_{s0}$ and measured $A_1^\square$ of magic-angle graphene follow the similar trend as doping varies[7,41], but systematic superfluid measurements are required to obtain a quantitative relation. In spite of large error bars, the data of iron-pnictide superconductors BFCA reside in the linear trend. For overdoped Tl2201, $\rho_{s0}$ and $A_1^\square$ were found to decrease in tandem with the increase of hole doping, as pointed out by Philips *et al.*[40], which is similar to our finding and implies a linear relation despite the relatively scarce data points. For overdoped LSCO, although the strong correlation between $\rho_{s0}$ and $A_1^\square$ still persists, a quantitative relationship cannot be pinned down due to the large uncertainty of the $A_1^\square$ data (see Supplementary Section 12). The data points of $A_1^\square$ and $\rho_{s0}$ are well fitted by a linear relation within error bar, particularly in the regime of $T_c < 12$K. To summarize, we plot the $\rho_{s0}/(\rho_{s0})^{max}$ versus $\widetilde{A}_1^\square/(\widetilde{A}_1^\square)^{max}$ relationship in Fig. 4a (the superscripts max refer to the maximum values in the literature), which suggests a linear scaling behavior that captures the common relation between $\rho_{s0}$ and $A_1^\square$ among different superconducting systems.

The similarity of ion-gated FeSe and overdoped cuprates revealed by the $\rho_{s0}$ versus $A_1^\square$ relation is surprising because they are distinct in many aspects. In overdoped cuprates, the normal-state carrier density exhibits an anticorrelation with $\rho_{s0}$ as the hole doping varies[13,18]. Whereas in FeSe, the ionic-liquid gating gives rise to the enhancement in $\rho_{s0}$, accompanied with an increase in the electron density and a decrease in the hole density (Supplementary Section 2). FeSe's nodeless gap (see e.g., Ref. [28] and Fig. 2b) also indicates that it differs from the cuprates in the pairing symmetry. Hence, the universal scaling shown in Fig. 4a indicates a fundamental link between superconducting condensation and the strange-metal scattering in unconventional superconductors. Since $\rho_{s0}$ is a crucial ingredient determining $T_c$ in unconventional superconductors[18-20,37], it may also provide a natural explanation of the scaling relation between $A_1^{\square}$ and $T_c$ ubiquitously observed during the last decades (see e.g., Refs. [1,2,9,40]).

To gain further insights into the correlation between $\rho_{s0}$ versus $A_1$, we perform the calculations based on a universal theory of strange metals, captured by a model involving metals of fermions coupled to quantum critical order-parameter scalars, dubbed as the 2d-YSYK model. The recent work unambiguously shows that the 2d-YSYK model[8,42,43] with spatially random Yukawa couplings exhibits the $T$-linear resistivity of strange metals, and can display the instability from the strange metal to the superconductivity[42] (see details in Supplementary Section 14). Strikingly, by varying the renormalized boson mass $M$ and the potential disorder stimulatingly, a linear $\rho_{s0}$ versus $A_1$ relation is reproduced (Fig. 4b). Here, $M$ is changed to tune the distance to the putative quantum critical point (QCP)[42], which is a basic effect of doping or gating in experiments. Notably, such a choice of model parameters is consistent with the fact that doping or gating also inevitably alters the amount of disorder, as evidenced by the evolution of the residual resistivity ratio (Supplementary Section 13) and was highlighted by recent studies of overdoped cuprates (see e.g., Refs. [44,45]). Disorder is a strongly relevant perturbation to clean quantum-criticality, which can explain the applicability of the 2d-YSYK theory to relatively clean samples[42,46]. The agreement between the theoretical calculations and the experimental data thus suggests a comprehensive consideration of quantum criticality and disorder provides a promising route for deciphering the interplay between strange metallicity and superconductivity in unconventional superconductors. As a consequence, combining the theoretical calculation on 2d-YSYK model, our experimental findings bring us closer to a unified understanding of the normal-state strange metal and superconductivity in unconventional superconductors.

In addition, a non-zero extrapolation of $A_1^{\square}$ to $\rho_{s0} = 0$ $\mu m^{-2}$ or $T_c = 0$ K is seemingly ubiquitous in overdoped cuprates including LSCO and Bi2201 (Ref. [14]), which is witnessed from the difference of the boundaries of superconducting dome ($p_{sc}$) and the strange-metal phase ($p_{sm}$) in the $T$ versus doping phase diagram, as depicted in Supplementary Section 15. This may stem from the suppression of superconductivity between $p_{\mathrm{sc}}$ and $p_{\mathrm{sm}}$ due to pair breaking[47] or the presence of competing orders[48]. Besides, it was suggested that the multi-orbital effect in iron-based superconductors, which involves orbital-dependent quasi-particle spectral weights in the normal state and orbital-selective pairing in the superconducting state[49], could contribute to the residual $A_1^{\square}$ (Ref. [10]). Whether these scenarios could account for the residual $A_1^{\square}$ remains to be investigated.

**Acknowledgments**

The authors would like to thank Dung-Hai Lee and Yuji Matsuda for fruitful discussions. This work was supported by the National Key Research and Development Program of China (Grants Nos. 2021YFA0718700, and 2022YFA1603903), the National Natural Science Foundation of China (Grants Nos. 12225412, 11927808, and 12274439), the CAS Project for Young Scientists in Basic Research (Grant No. 2022YSBR-048), the Key-Area Research and Development Program of Guangdong Province (Grant No. 2020B0101340002), the Beijing Nova Program of Science and Technology (Grant No. 20220484014).

**Author contributions**

K.J. and Z.X.Z. conceived the project. M.Y.Q. and Z.X.L. supervised the project. M.Y.Q. and R.Z.Z. designed the two-coil mutual inductance measurement device. R.Z.Z and M.Y.Q developed the fast wavelet collocation method for extracting the magnetic penetration depth from the pickup coil voltage. R.Z.Z. and M.Y.Q. performed the ionic liquid gating and the superfluid density measurements, with help from J.X., W.X.C., X.W.W. and Q.Y.S.. C.Y.L. performed the calculations based on the two-dimensional YSYK model, with the guidance of S. S.. Z.Y.Z. synthesized the FeSe films. M.Y.Q. and R.Z.Z. analyzed the experimental data, with assistance from Z.X.W., X.Y.J., J.Y., Y.M.L., and Q.H.C.. T.X., Z.X.L. and S.S. contributed to the theoretical discussions. R.Z.Z., M.Y.Q., C.Y.L., S.S., Z.X.L. and K.J. wrote the manuscript with input from all authors.

**Competing interests**

The authors declare no competing interests.

**Data availability**

Source data are provided with this paper. All other data that support the findings of this study are available from the corresponding authors upon reasonable request.

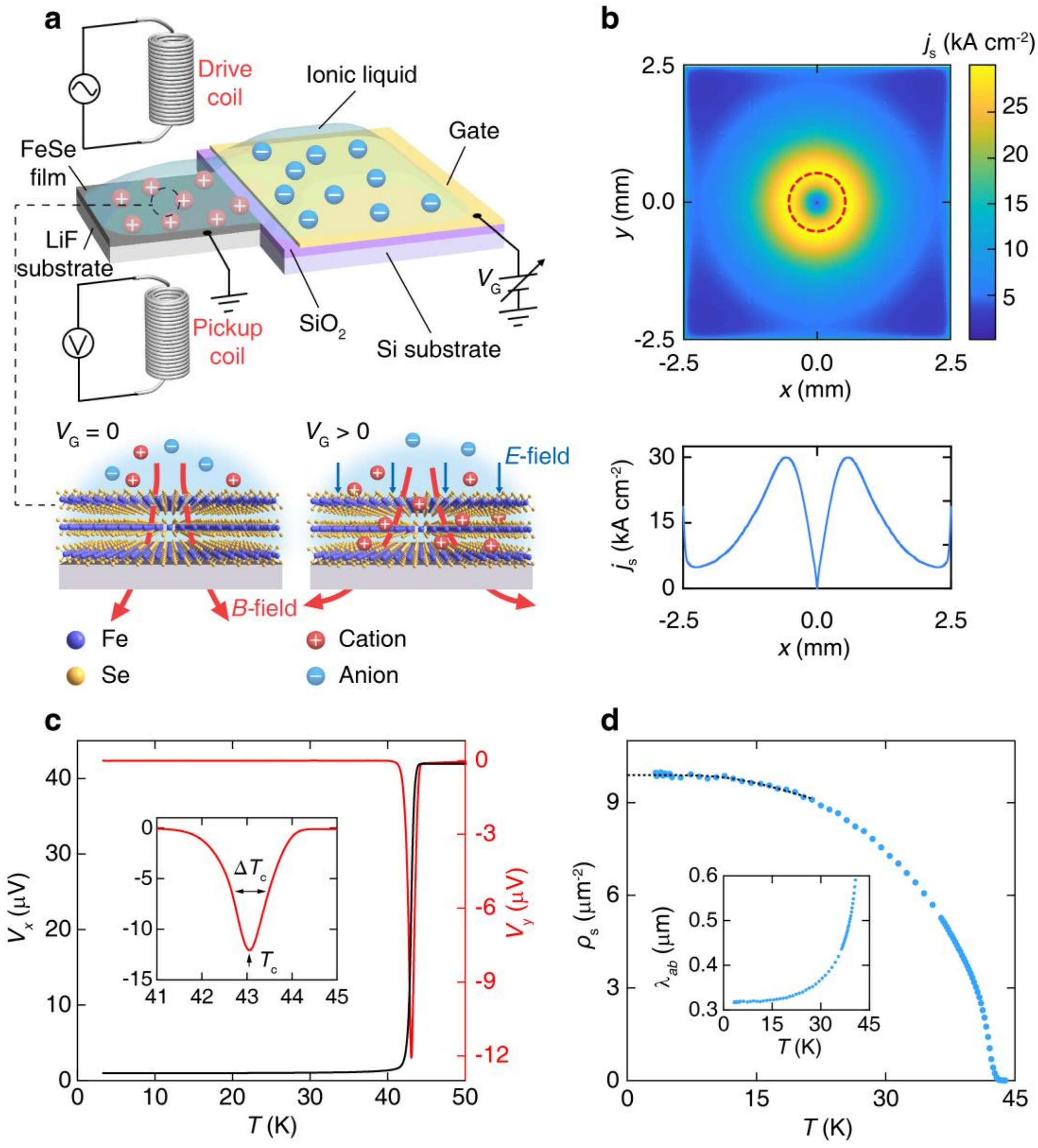

**Fig. 1 | Operando superfluid density measurements on the ion-gated FeSe film. a,** Schematic of a typical ILG device and the TCMI measurement scheme. The ILG device consists of a FeSe film and a gate electrode, with the ionic liquid DEME-TFSI covered. A positive gate voltage ($V_G$) is applied between the gate electrode and the film to induce the electrochemical protonation. A drive coil and a pickup coil are mounted on opposite sides of the ILG device for in-situ measurements of the diamagnetic response of the sample. The lower picture shows the enlarged schematic of the ion-gated FeSe. Under a positive $V_G$, the cations in the DEME-TFSI will enter the FeSe film, resulting in an enhancement of the superfluid density. Consequently, the diamagnetic response becomes stronger, from which the evolution of superfluid density with ionic-liquid gating can be extracted. **b**, Top view of the induced screening current density, $j_s$, for the

ion-gated FeSe film with $T_c \approx 43$ K, calculated using the FWC method[25] (Supplementary Section 4). The red dashed line shows the position of the drive coil. The lower panel shows the cut at $y = 0$. **c,** Temperature dependence of the pickup coil voltage, $V = V_x + iV_y$, for a gated state with $T_c \approx 43$ K. The sudden drop of $V_x$ (black line) around $T_c$ reflects the strong diamagnetism of the superconducting state. Accordingly, $V_y$ (red line) shows a dip and its full width at half maximum is smaller than 1 K, indicating good homogeneity of superconductivity[20]. Inset: Zoom-in of the $V_y$ versus $T$ curve. $T_c$ and $\Delta T_c$ are the position and the full width at half maximum of the dip, respectively, following the definitions in Ref. [24]. **d,** Temperature-dependent superfluid density, $\rho_s(T)$, extracted from the pickup coil voltage in **c**. The dashed line represents the nodeless-gap fit (Supplementary Section 6). Inset: Temperature-dependent in-plane magnetic penetration depth, $\lambda_{ab}(T)$.

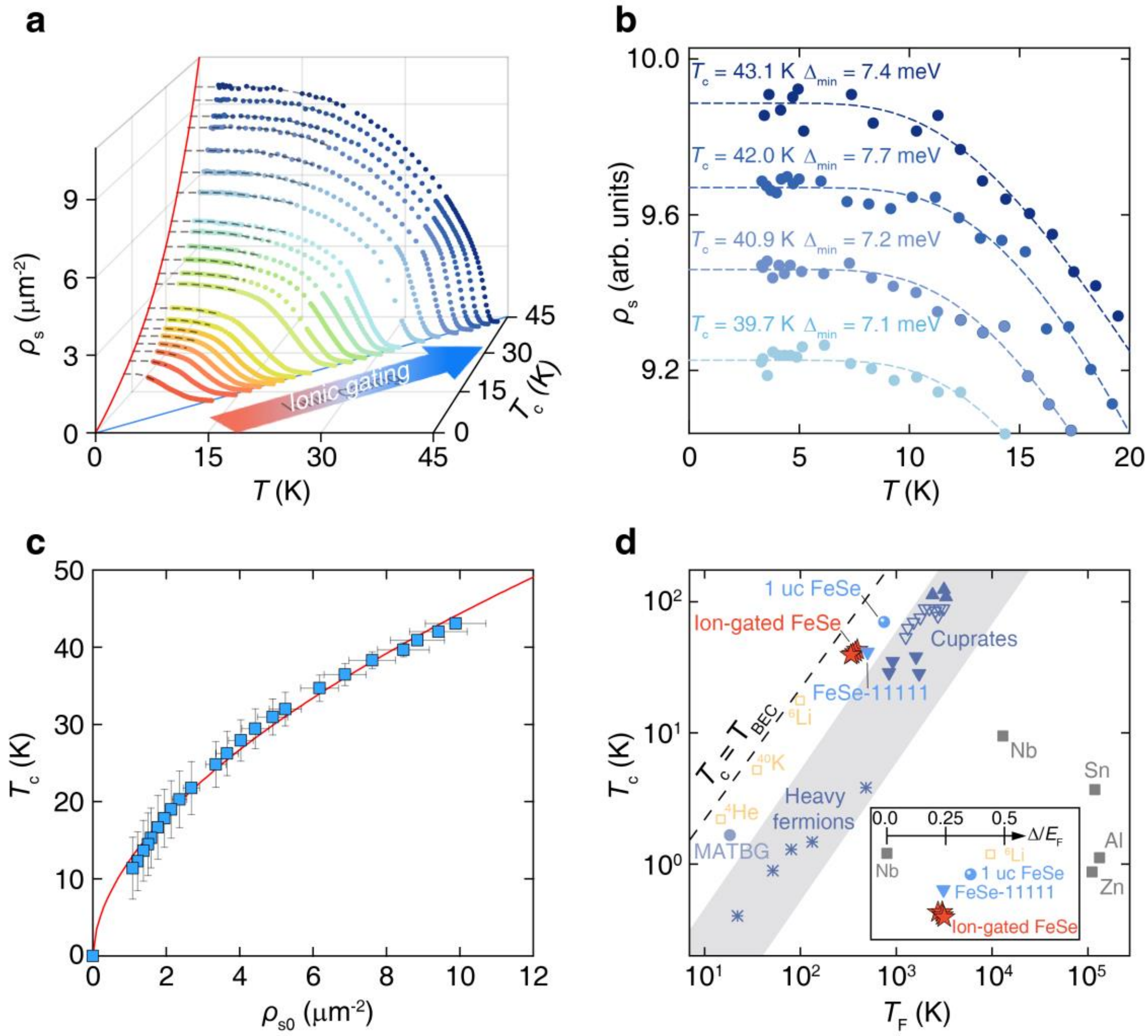

**Fig. 2 | Evolution of the superfluid density in the ion-gated FeSe film. a,** Temperature-dependent superfluid density, $\rho_s(T)$, for the ion-gated FeSe film with $T_c$ systematically tuned from approximately 11 to 43 K. The dashed lines are the nodeless-gap fits to the first 10% drop in $\rho_s(T) / \rho_s(T \rightarrow 0)$, as depicted in the expanded view of the $\rho_s$ data at low temperatures in **b**. **c,** Dependence of $T_c$ on the zero-temperature superfluid density, $\rho_{s0} \equiv \rho_s (T \rightarrow 0)$, where $\rho_{s0}$ is determined from the nodeless-gap fit. The definition of $T_c$ is shown in Fig. 1c. The blue squares represent the experimental data; the red solid line is the fit to $T_c = \gamma \rho_{s0}^{0.55}$, where $\gamma$ is the fit parameter. **d,** Uemura plot: $T_c$ versus Fermi temperature, $T_F$, for various superconductors[50]. The ion-gated FeSe with $T_c > 39$ K, as well as other FeSe-based superconductors, i.e., monolayer FeSe on $SrTiO_3$ substrate (1 uc FeSe) and $(Li_{0.8}Fe_{0.2})OHFeSe$ (FeSe-11111) are located closer to the BEC limit for three-dimensional bosonic gas ($T_c = 0.218T_F$, the dashed line) than cuprate and heavy-fermion superconductors (the shade region). The large ratio of $\Delta/E_F$

≈ 0.23 for ion-gated FeSe also indicates it lies in the BCS-BEC crossover regime (see the inset). $T_F$ and $E_F$ are obtained from $\rho_{s0}$ using the formula given in Ref. [28], i.e., $E_F = k_B T_F = \frac{\pi \hbar^2 d}{\mu_0 e^2} \rho_{s0}$, where $d$ is the interlayer distance. The values of $\Delta$ are obtained from the nodeless-gap fits in **b**.

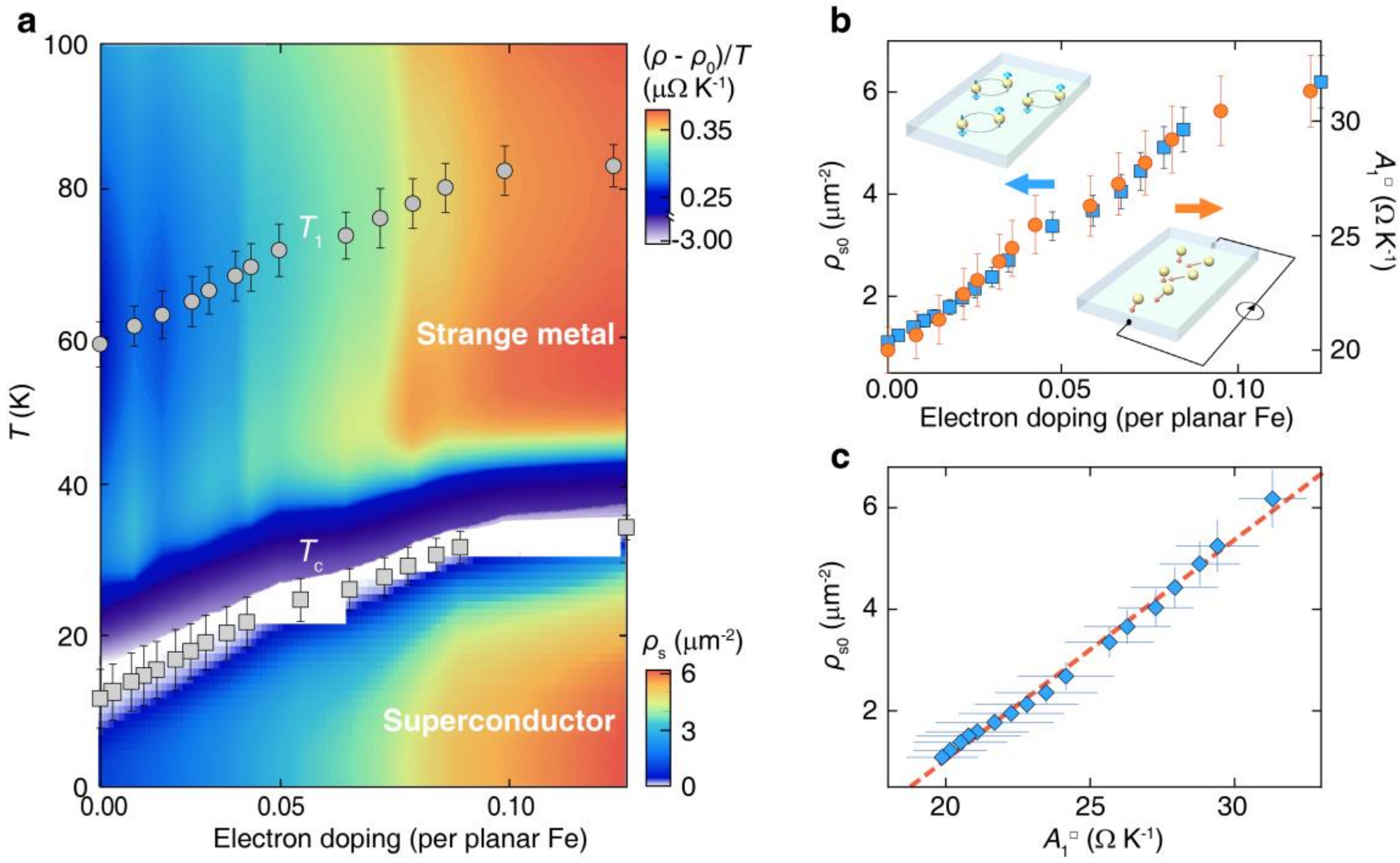

**Fig. 3 | Correlation between unconventional superconductivity and the strange metallicity in the ion-gated FeSe film. a,** Synchronized evolution of the strange-metal phase and the superfluid condensate with ionic-liquid gating. The strength of the strange-metal scattering, $(\rho - \rho_0)/T$, and the magnitude of the superfluid density, $\rho_s \equiv \lambda_{ab}^{-2}$, versus temperature $T$ and election doping level are shown by the contour plot. The gray circles and gray squares represent the upper bound temperature of $T$-linear resistivity ($T_1$) and $T_c$, respectively. It is shown that the strange-metal phase [characterized by $(\rho - \rho_0)/T$ and $T_1$] and the superfluid condensate (characterized by $\rho_s$ and $T_c$) changes synchronously with doping. Here $\rho_0$ is the residual resistivity obtained by fitting the resistivity at low temperatures (see the inset of Fig. S7) with the function $\rho = \rho_0 + A_1 T$. The normal state data are extracted from Fig. S8. The doping levels are extracted from the $T_c$ versus doping relation in Ref. [10]. **b,** The zero-temperature superfluid density, $\rho_{s0}$, (blue squares) and the $T$-linear coefficient per Fe-Se layer, $A_1^{\square}$, (red circles) versus electron doping in the ion-gated FeSe film. Here $A_1^{\square} = A_1/d_{\text{Fe-Se}}$, where $d_{\text{Fe-Se}}$ is the distance between adjacent Fe-Se layers. Evidently, $\rho_{s0}$ tracks $A_1^{\square}$ closely with electron doping, demonstrating a direct link between the formation of the superfluid in the ground state and the scattering or carriers in the strange-metal normal state. **c**, $\rho_{s0}$ as a function of $A_1^{\square}$ in the ion-gated FeSe film.

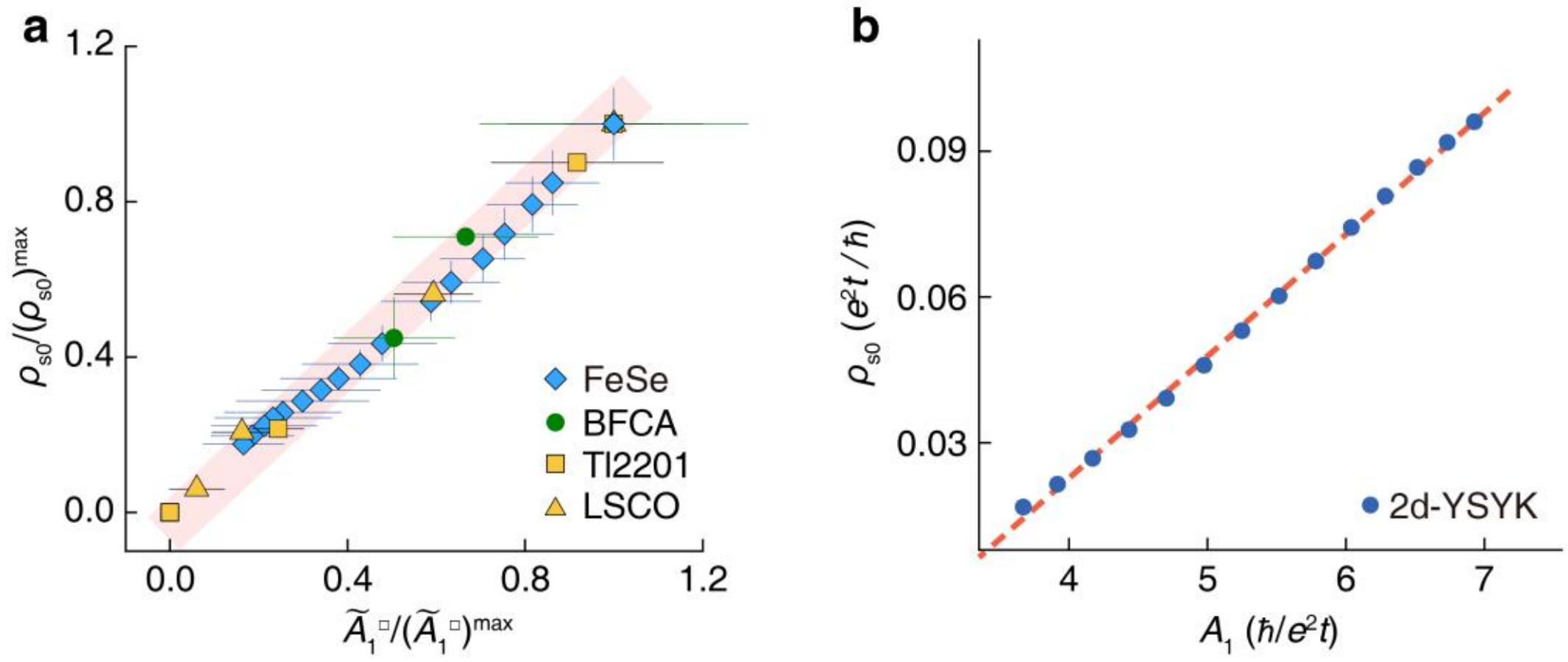

**Fig. 4 | Scaling of the zero-temperature superfluid density and the *T*-linear resistivity coefficient for unconventional superconductors. a,** Scaling of $\rho_{s0}$ with $\widetilde{A}_1^{\square}$ for different superconducting systems, where $\rho_{s0}$ and $\widetilde{A}_1^{\square}$ are normalized by their respective maximum values in the literature. Here, a constant has been subtracted from $A_1^{\square}$, namely $\widetilde{A}_1^{\square} = A_1^{\square} - A_0$, where $A_0$ represents the non-zero extrapolation of $A_1$ to $T_c$ = 0 K (Ref. [10]). Data for $Ba(Fe_{1-x}Co_x)_2As_2$ (BFCA) are extracted from Refs. [9,30]. Data for $Tl_2Ba_2CuO_{6+\delta}$ (Tl2201) are obtained from Ref. [40] and references therein. Data for heavily overdoped $La_{2-x}Sr_xCuO_4$ (LSCO, $T_c \leq 12$ K) are extracted from Refs. [10,18]. The error bars are reproduced from published data. The red shading is a guide to the eye. **b,** $\rho_{s0}$ versus $A_1$ relation calculated based on the two-dimensional Yukawa-Sachdev-Ye-Kitaev (2d-YSYK) model (Supplementary Section 14), where $\hbar$ is the reduced Planck constant and $t$ is the fermion hopping.

# Supplementary Information

## 1. Film characterizations

To characterize the structural property of the film, we have performed the x-ray diffraction (XRD) measurements on FeSe films with two Ge (220) single crystals, as shown in Fig. S1. The XRD $\theta$-$2\theta$ scan pattern shows a high-quality (00*l*) oriented growth (Fig. S1a), i.e., the *ab* plane is parallel to the film surface. The full width at half maximum (FWHM) of the XRD rocking curve is 0.477°, showing high crystalline quality (Fig. S1b). In addition, the high-quality epitaxy of the film is confirmed by a clear four-fold symmetry in the XRD $\varphi$ scan pattern for the (011) diffraction peak (Fig. S1c). The *c*-axis lattice parameter calculated from the XRD $\theta$-$2\theta$ scan data by Bragg's law is 5.54 Å, which is close to the 5.48 Å of bulk FeSe (Ref. 1).

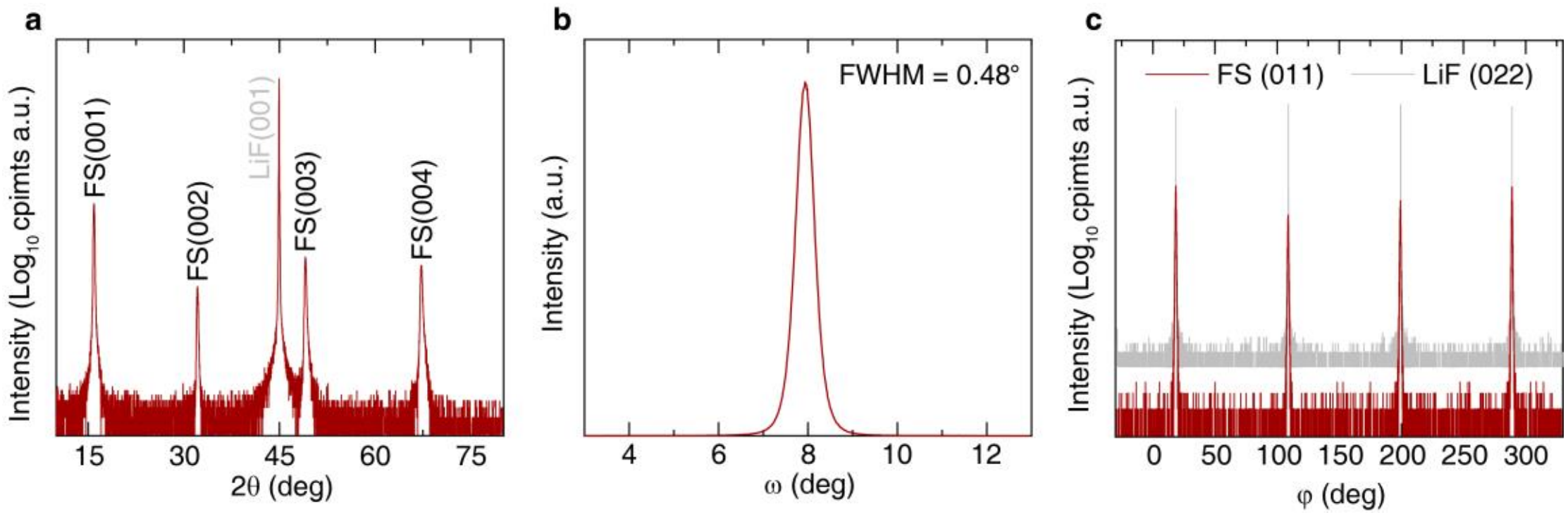

**Fig. S1. XRD characterizations of a typical FeSe (FS) film grown on LiF substrate**. **a**, The $\theta$-$2\theta$ diffraction pattern. Only (00*l*) peaks are observed for FeSe, indicating high crystallinity along the *c*-axis. **b**, The rocking curve of (001) peak, showing a narrow full width at half maximum (FWHM) of 0.48°. **c**, The $\varphi$-scan result of FeSe (011) and LiF (022) peaks, both of which show clear four-fold symmetry.

The chemical composition distributions of the film were characterized by the energy-dispersive x-ray (EDX) spectroscopy with a Hitachi SU5000 field-emission scanning electron microscope (SEM). Figure S2a shows a typical SEM micrograph of the film, showing a relatively flat surface. The averaged EDX spectrum of the image area is depicted in Fig. S2b. The EDX mappings of the image area are shown in Fig. S2c-d, demonstrating no signature of inhomogeneous compositional distributions on the micron scale.

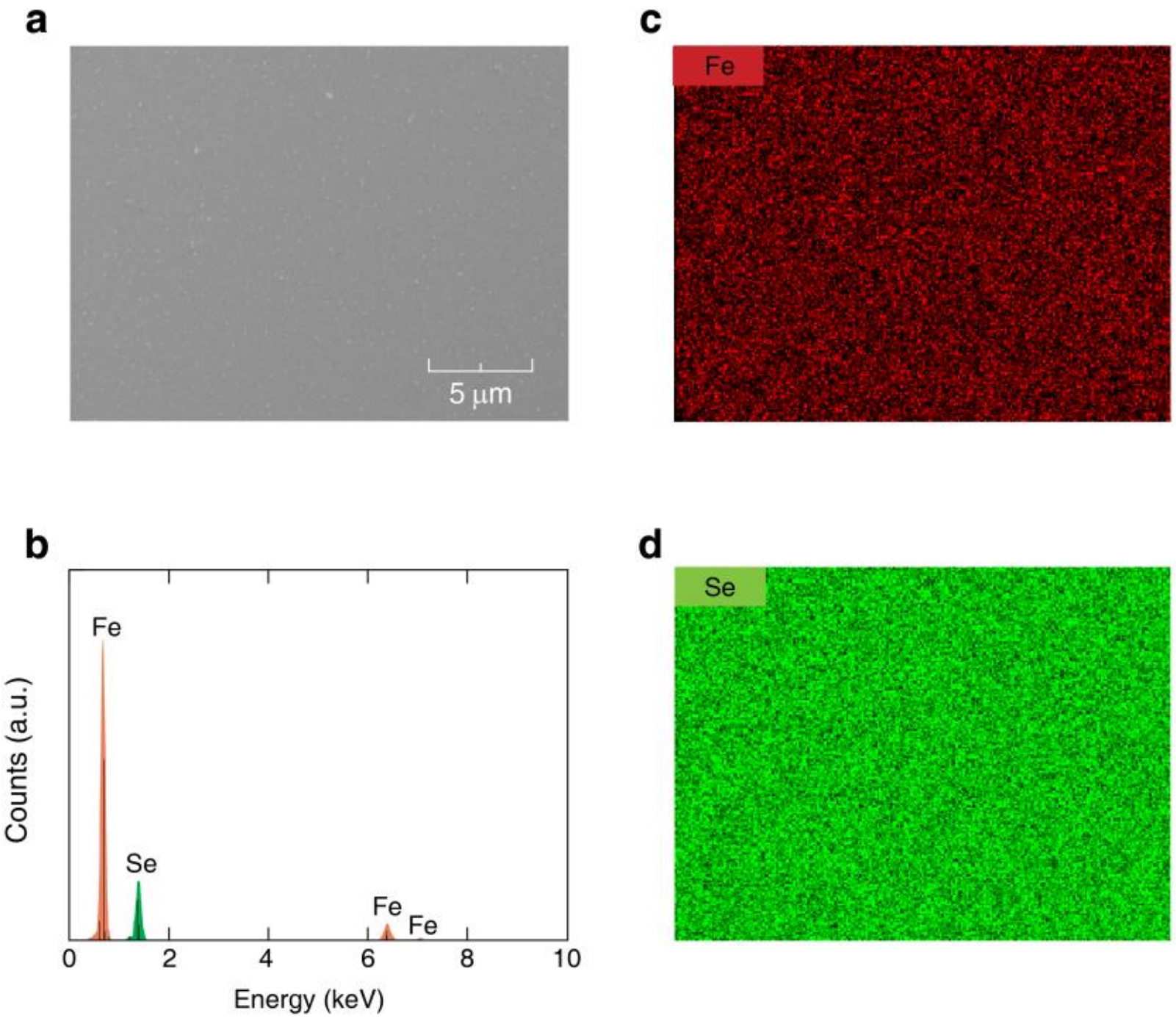

**Fig. S2. EDX characterizations of a typical FeSe (FS) film grown on LiF substrate**. **a**, SEM micrograph. **b**, EDX spectra for chemical compositions. **c-d**, EDX mappings corresponding to the SEM imagine area in **a**.

## 2. Evolution of the normal-state carrier density with ionic-liquid gating

FeSe is a multi-band system with hole pockets at Γ point and electron pockets at M point (see e.g., Ref. 2). With ionic-liquid gating, it was found that the hole density decreases and the electron density increases[3,4]. Meanwhile, the superfluid density increases, as depicted in Fig. S3.

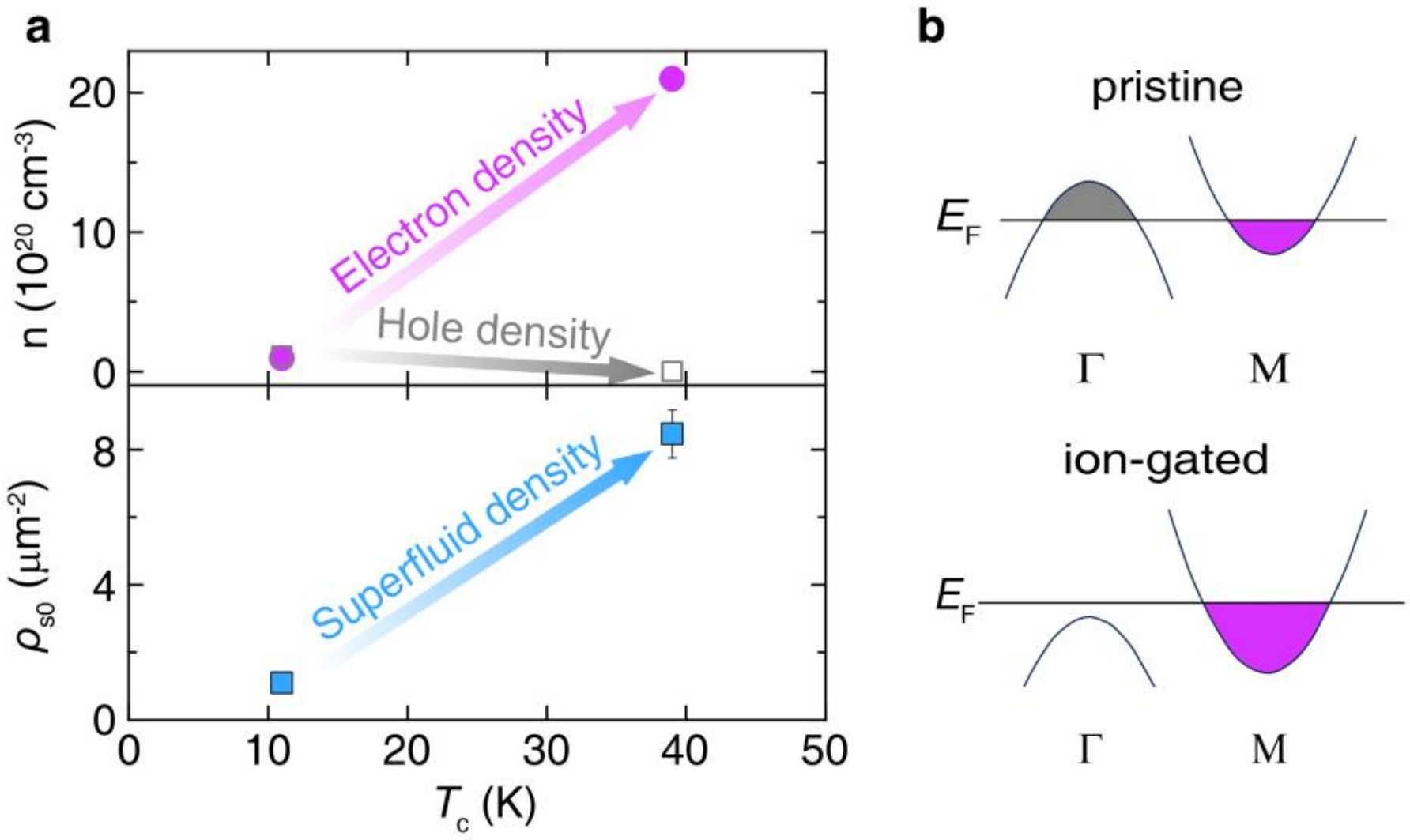

**Fig. S3. a**, Evolution of the normal-state carrier density and the superfluid density with ionic-liquid gating. Data of the normal-state carrier density are obtained from Table. 1 in Ref. 4. **b**, Schematic illustrations of the band structure before (top) and after (down) ionic-liquid gating.

### 3. Device configuration and the in-situ TCMI measurement

Figure S4a shows a photograph of the ILG device. A FeSe/LiF film with size of $5 \times 5\ \mathrm{mm}^2$ is placed next to a Si substrate covered by the $SiO_2$ insulating layer. A gate electrode with size of $10 \times 7\ \mathrm{mm}^2$ is fabricated on the $SiO_2$ layer by depositing Au/Ti (50 nm/5 nm) with electron-beam evaporation. Electrical contacts are made by indium (In) soldering on the side of the FeSe film and a corner of the gate electrode. The ionic liquid DEME-TFSI is covered by a 6-μm-thick Kapton foil to confine its coverage on the sample and the gate electrode.

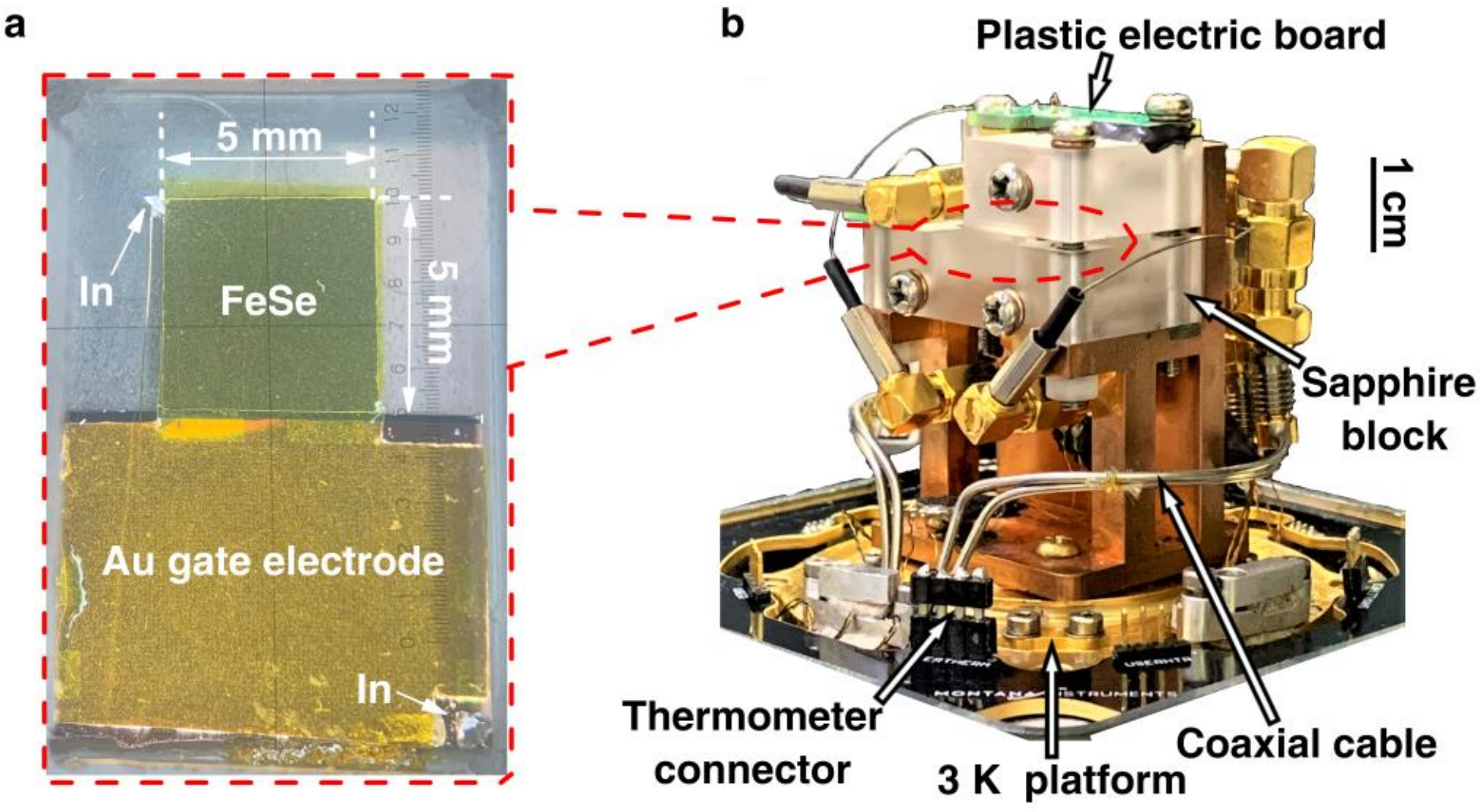

**Fig. S4. Photographs of a, the ILG device and b, the TCMI apparatus**. The ILG device is mounted inside the TCMI apparatus, indicated by the dashed line.

During the ionic-liquid gating, the ILG device is mounted inside a sapphire block of the TCMI apparatus (Fig. S4b). A drive coil and a pickup coil are sealed in the sapphire block with epoxy and aligned axially with the center of the FeSe film. The two coils are the same in size: The number of turns is 300, the inner diameter is 0.5 mm, the outer diameter is 1.3 mm, and the length is 1.6 mm. The pickup coil is pressed against the backside of the LiF substrate, while the drive coil is about 0.4 mm away from the surface of the FeSe film. A plastic electric board is fastened on the sapphire block to provide a platform for wiring. To reduce noise and interference, coils are connected to coaxial cables. The apparatus is thermally connected to the 3 K platform of a Montana Instruments cryocooler. A Lakeshore thermometer is mounted inside the sapphire block to measure the sample temperature. A Keithley 2450 source meter is used to apply the gate voltage and monitor the leakage current. A Stanford Research SR830 lock-in amplifier is used to supply the alternating current to the drive coil. The current amplitude is 0.2 mA, and the frequency is 50 kHz, which is in the linear-response regime[5]. The induced pickup coil voltage, $V = V_x + iV_y$, is probed by the same lock-in amplifier with a reference phase of 90°.

A complete experiment includes a set of sequences, each of which involves 1) warming up to a target temperature ($T_G$) under a positive gate voltage ($V_G$), 2) staying at the target temperature for a given period of time ($t_G$), and 3) cooling down to measure the induced pickup coil voltage.

## 4. Extracting the magnetic penetration depth using the FWC method

Among various techniques for measuring the magnetic penetration depth[6] ($\lambda_{ab}$), the TCMI technique has attracted increasing attention due to its high sensitivity, simplicity, and flexibility[7-12]. A transmission-type TCMI system consists of a drive coil and a pickup coil, which are located on opposite sides of the superconducting film, as shown in Fig. 1a of the main text. A magnetic field is produced by the alternating current in the drive coil, inducing a voltage, $V = V_x + iV_y$, in the pickup coil. The real component, $V_x$, represents the inductive coupling, whereas the imaginary component, $V_y$, represents the resistive coupling.

When the films enter the Meissner state, two types of currents in the system will contribute to $V_x$. One is the alternating current applied to the drive coil, whose contribution can be calculated with the coil parameters based on classical electrodynamics. The other is the screening current induced in the superconducting film, which is related to $\lambda_{ab}$ through[13]

$$\mathbf{j}_s(\mathbf{r}) + \frac{\sinh(d/\lambda_{ab})}{4\pi\lambda_{ab}} \int_{\Omega} d^2\mathbf{r}' \frac{\mathbf{j}_s(\mathbf{r}')}{|\mathbf{r}-\mathbf{r}'|} = -\frac{1}{\mu_0 {\lambda_{ab}}^2} \mathbf{A}_d(\mathbf{r}), \quad \text{(S1)}$$

where $\mathbf{j}_s$ is the screening current density, $\Omega$ is the projection of the film on the $x$-$y$ plane (film surface), $\mathbf{r} = \hat{\mathbf{x}}x + \hat{\mathbf{y}}y$, $d$ is the film thickness, $\mu_0 = 4\pi \times 10^{-7}$ N A$^{-2}$ is the vacuum permeability, and $\mathbf{A}_d$ is the vector potential generated by the drive current. Employing the FWC algorithm, Eq. S1 can be accurately solved and $\mathbf{j}_s$ is obtained (Fig. 1b in the main text). Then the pickup coil voltage, $V_{cal}$, can be calculated by integrating the vector potential of both the drive and screening currents around each loop of the pickup coil. Finally, the absolute $\lambda_{ab}$ can be extracted from a lookup table consisting of $V_{cal}/V_{cal}(T > T_c)$ for different $\lambda_{ab}$ values. More details about this procedure can be found in Ref. 13.

## 5. Evolution of the diamagnetic responses in the ion-gated FeSe film

Figure S5 shows the raw TCMI data used for determining the superfluid density in Fig. 2a of the main text. With ionic-liquid gating, $T_c$ of the FeSe film is systematically tuned from approximately 8 to 43 K, reflecting as a continuous shift of the dip of the Vy(T ) curve. For each gated state, only a single magnetic transition is observed, confirming the bulk modulation of the superconductivity.

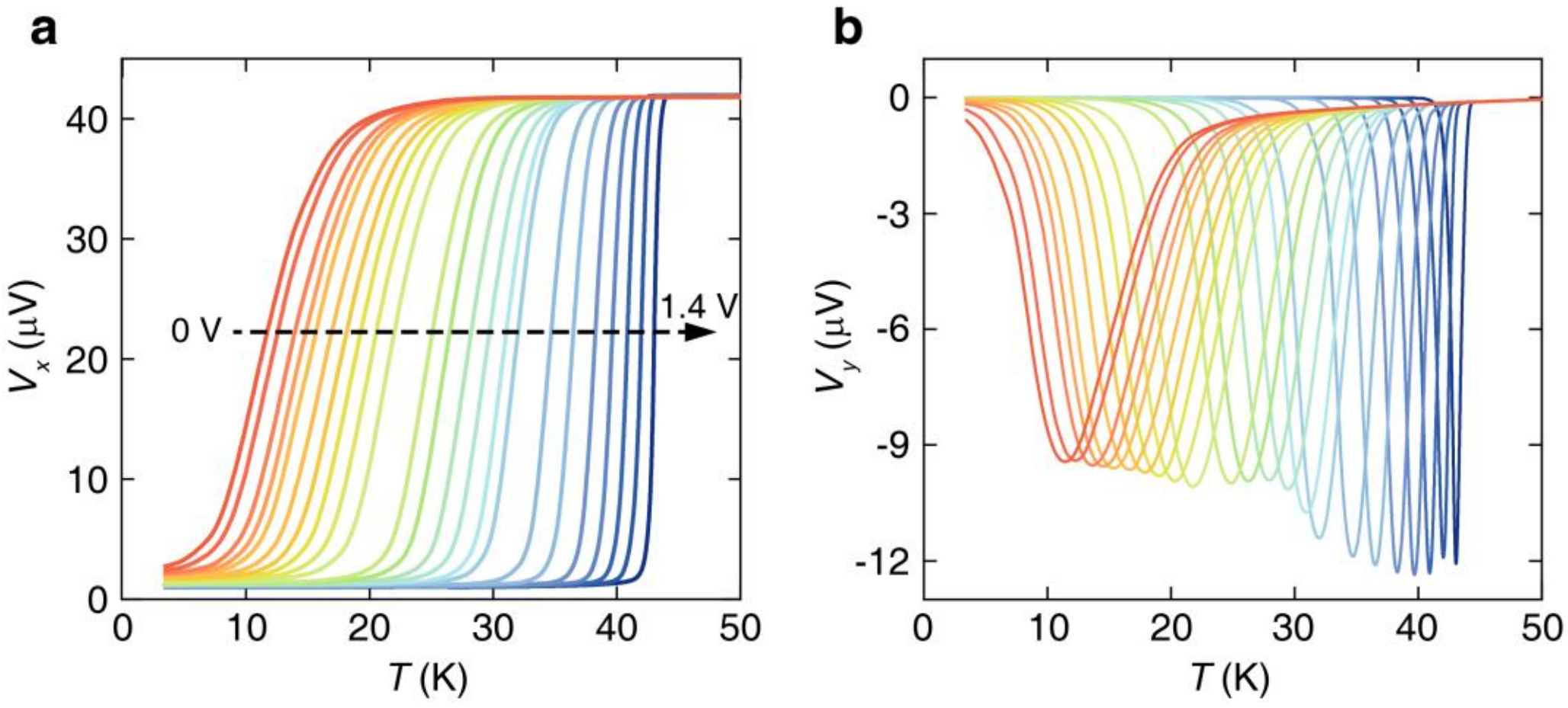

**Fig. S5. Evolution of the diamagnetic responses in the ion-gated FeSe film. a,** The real and **b,** the imaginary components of the pickup coil voltage, $V = V_x + iV_y$, as a function of temperature for successive gated states, respectively. From left to right, the gate voltages are $V_G$ = 0, 0.69, 0.7, 0.71, 0.73, 0.75, 0.76, 0.77, 0.78, 0.79, 0.8, 0.83,

0.87, 0.91, 0.92, 0.97, 0.99, 1.06, 1.12, 1.18, 1.2, 1.4 V. $T_G \approx 325$ K and $T_G \approx 3$ h for each curve.

## 6. Determining $\rho_{s0}$ based on different gap symmetries

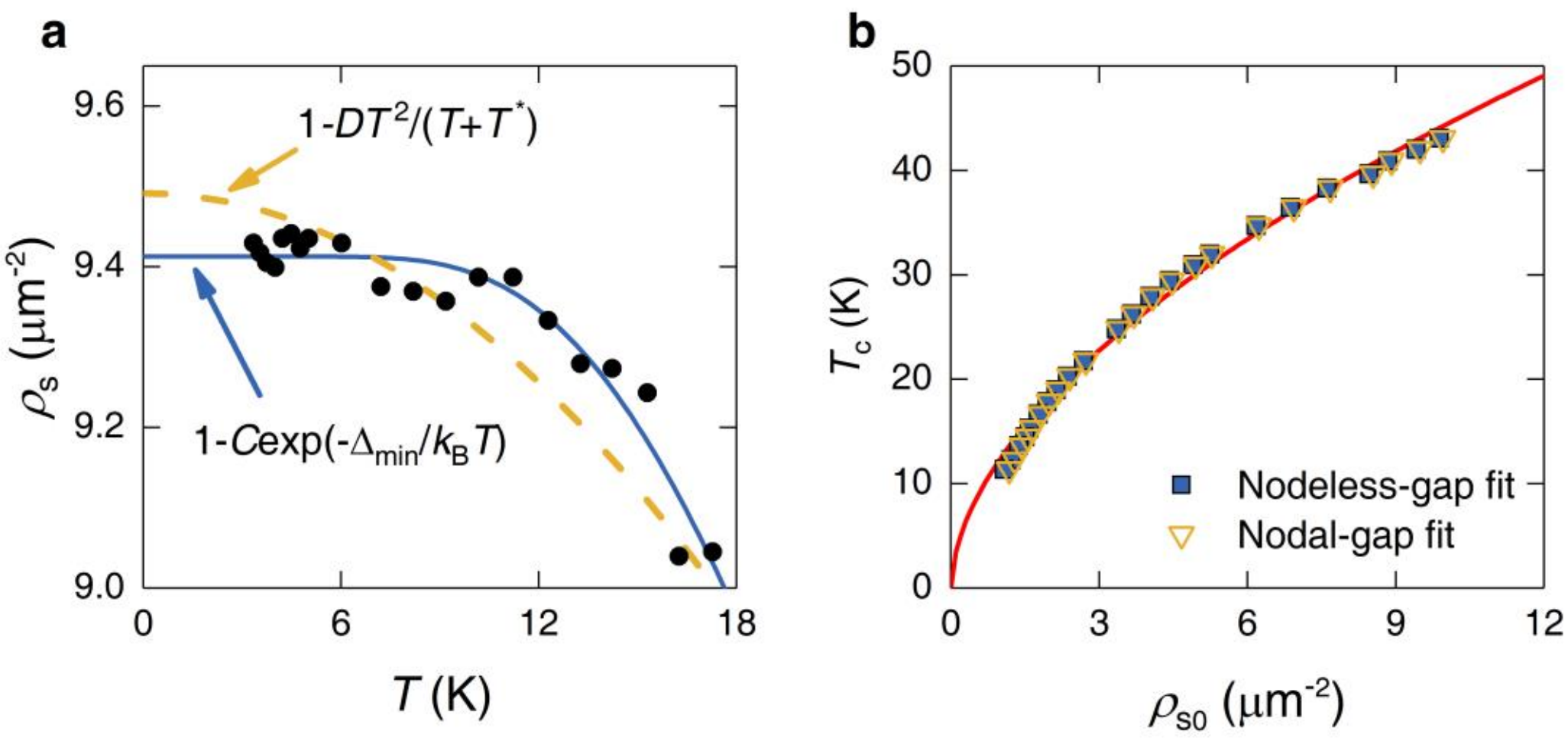

**Fig. S6. Determination of the zero-temperature superfluid density, $\rho_{s0} \equiv \rho_s(T \rightarrow 0)$. a,** Expanded view of the $\rho_s$ data at low temperatures for the sample with $T_c \approx 42$ K. The solid and dashed lines denote the best fits using Eq. S2a and S2b to the first 10% drop of the $\rho_s(T)$ data, respectively. **b,** Dependence of $T_c$ on $\rho_{s0}$, where $\rho_{s0}$ is extrapolated using Eq. S2a (blue solid squares) and Eq. S2b (brown hollow triangles), respectively. The solid line indicates $T_c = \gamma \rho_{s0}^{0.55}$, as depicted in the main text.

To determine the extrapolated value of the zero-temperature superfluid density, $\rho_{s0} \equiv \rho_s(T \rightarrow 0)$, we have fitted the first 10% drop of the $\rho_s(T)$ data using the equations for nodeless-gap superconductors[14]

$$\rho_s(T)/\rho_{s0} = 1 - C\exp(-\Delta_{min}/k_B T), \tag{S2a}$$

and nodal-gap superconductors[9,14,15]

$$\rho_s(T)/\rho_{s0} = 1 - D\frac{T^2}{T+T^*}, \tag{S2b}$$

respectively. Here, $\Delta_{min}$ is the minimum gap size at $T = 0$ K, $k_B$ is the Boltzmann constant, $C$ and $D$ are fit parameters, and $T^*$ represents the crossover temperature from quadratic to linear behavior. Figure S6a is an expanded view of the low-temperature behavior of $\rho_s$ for a gated state with $T_c \approx 42$ K. It is shown that the nodeless-gap fit (the blue solid line) is in better agreement with our data, implying a nodeless pairing gap in ion-gated FeSe. Nevertheless, the dependence of $T_c$ on $\rho_{s0}$ obtained with two gap symmetries are almost the same, as displayed in Fig. S6b.

## 7. Failure of the dirty BCS model to account for the $T_c$ versus $\rho_{s0}$ scaling

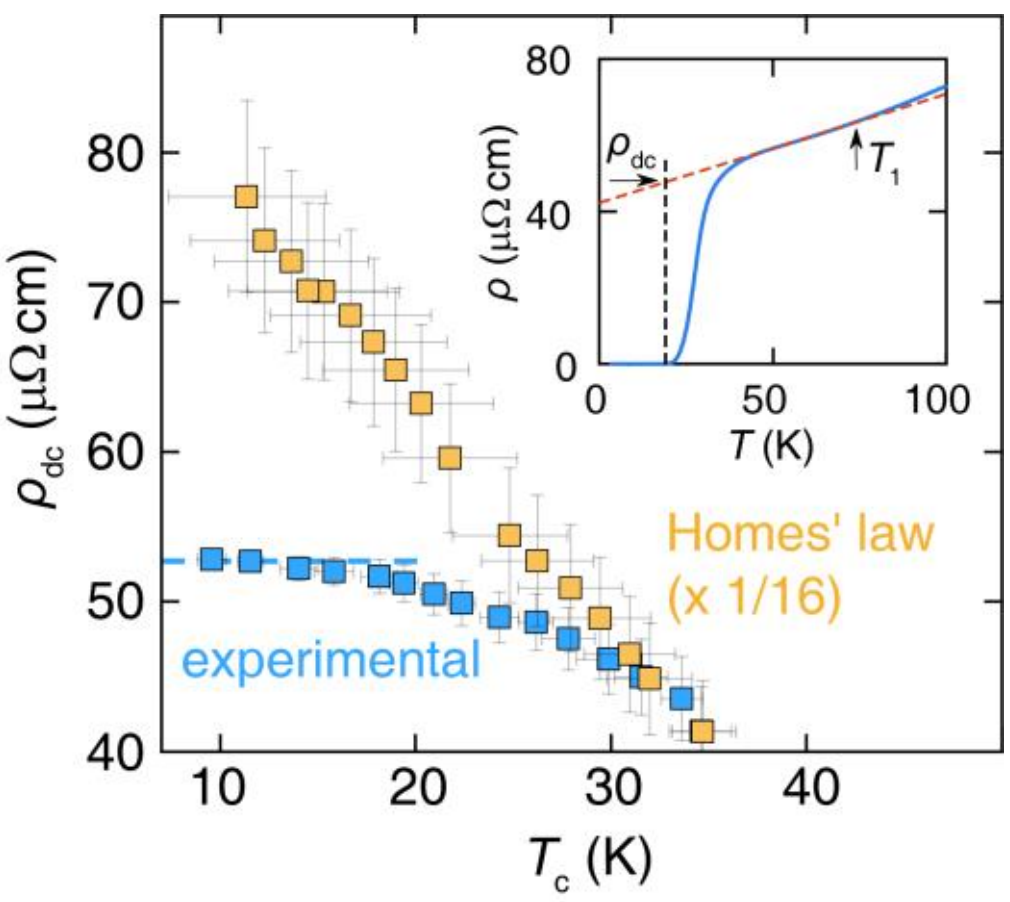

**Fig. S7. Comparison of $\rho_{dc}$ values calculated using Homes' law with measured values.** The orange squares represent $\rho_{dc}$ calculated from measured $T_c$ and $\rho_{s0}$ values by applying Homes' law (multiplied by 1/16), i.e., $\rho_{s0} \propto \sigma_{dc} T_c$, which follows from the Ferrell-Glover-Tinkham sum rule for dirty BCS superconductors. Blue squares are the $\rho_{dc}$ obtained by extrapolating the $T$-linear resistivity, i.e., $\rho = \rho_0 + A_1 T$, above superconducting transition (the red dashed line) to $T_c$, as depicted in the inset. $T_1$ represents the upper bound temperature of $T$-linear resistivity. It is clear that $\rho_{dc}$ calculated using Homes' law grows rapidly with $T_c$ decreases, whereas the measured $\rho_{dc}$ saturates when $T_c$ < 15 K (see the blue dashed line).

## 8. Evolution of the electrical transport properties in the ion-gated FeSe film

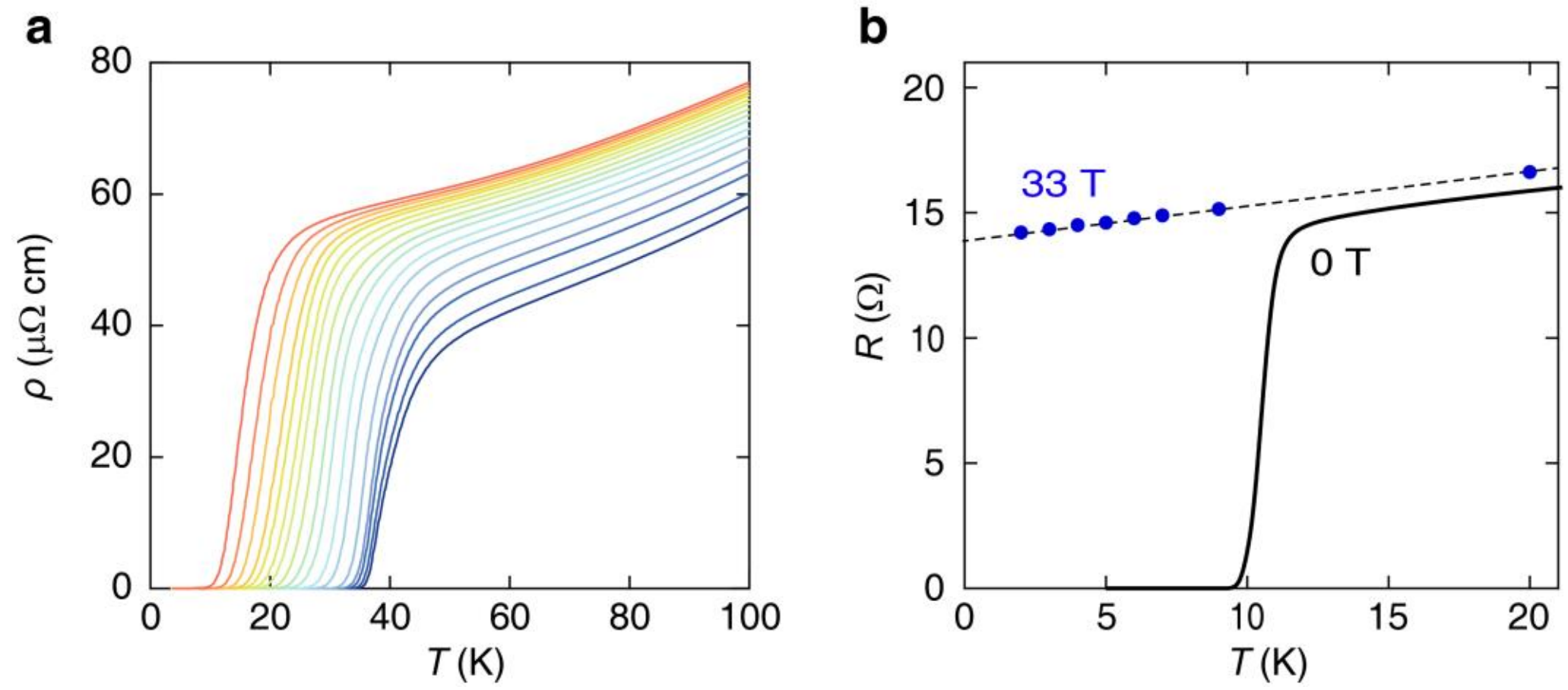

**Fig. S8. Evolution of the electrical transport in the ion-gated FeSe film. a,**

Temperature dependence of the resistivity for successive gating sequences. **b,** Temperature dependence of the resistance at zero field (solid line) and high field (33 T; blue dots). The black dashed line is a guide for $T$-linear resistivity. Data are extracted from our recent work[16].

## 9. Fitting the $\rho_{s0}$ versus $A_1^{\square}$ relation with error bars

By fitting the data points in Fig. 3c together with the error bars using a more general power-law formula, i.e., $A_1^{\square} \sim \rho_s^m$, the power-law exponent $m$ is found to be $0.978 \pm 0.063$, which is close to 1 and supports a linear relationship. We emphasize that the linear relationship is the most reasonable function to fit the data with the error bars taken into account. To further demonstrate this point, we present the fitting results using several common functions: root ($A_1^{\square} \sim \rho_{s0}^{1/2}$); linear ($A_1^{\square} \sim \rho_{s0}$); and quadratic ($A_1^{\square} \sim \rho_{s0}^{2}$) power-law dependences in Fig. S9. It is obvious that the data are in better agreement with the linear fit (Fig. S9b) than the root (Fig. S9a) and the quadratic (Fig. S9c) fits.

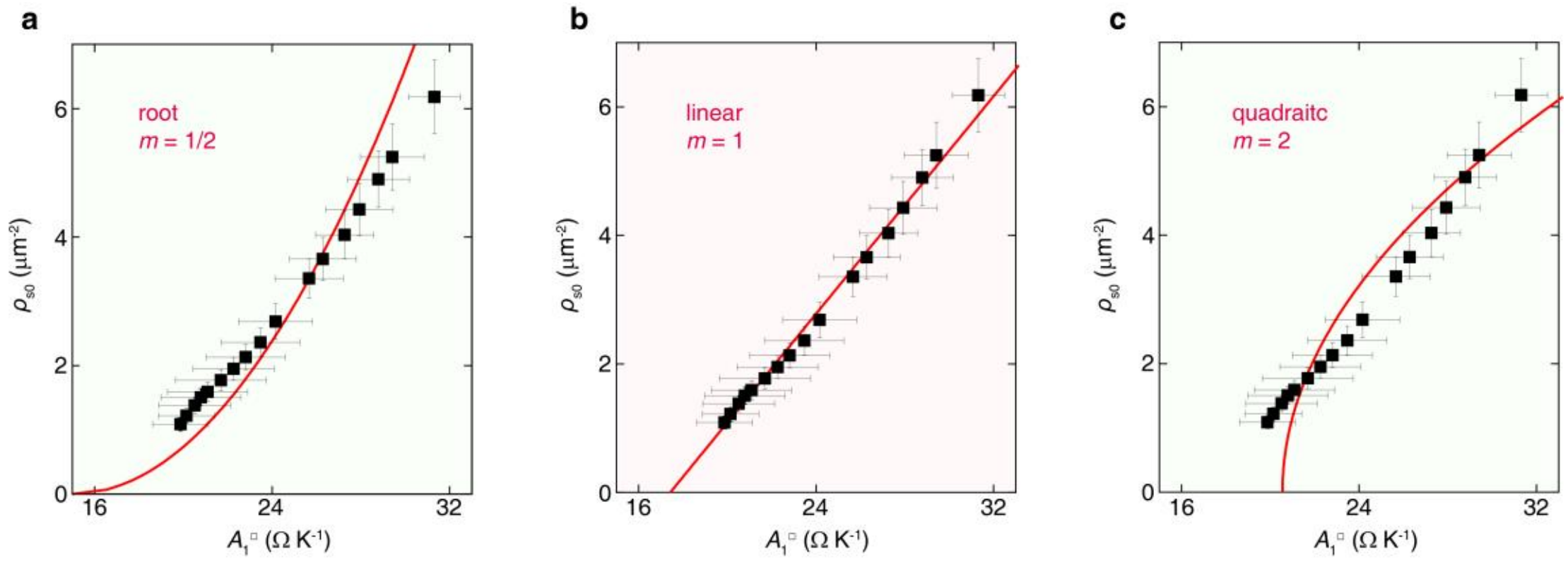

**Fig. S9. Best fits of the $\rho_{s0}$ versus $A_1^{\square}$ data together with the error bars in Fig. 3b to a, root ($A_1^{\square} \sim \rho_{s0}^{1/2}$); b, linear ($A_1^{\square} \sim \rho_{s0}$); and c, quadratic ($A_1^{\square} \sim \rho_{s0}^{2}$) power-law dependences.**

## 10. Raw data shown in Fig. 3c

**Table S1.** T-linear coefficients and zero-temperature superfluid density shown in Fig. 3c of the main text.

| $T_c$ (K) | $A_1^{\square}$ ($\Omega$ K$^{-1}$) | $\rho_{s0}$ ($\mu$m$^{-2}$) |
|---|---|---|
| 34.69 | 31.32 | 6.18 |
| 31.98 | 29.42 | 5.25 |
| 30.96 | 28.79 | 1.4 |
| 29.43 | 27.94 | 4.43 |
| 27.92 | 27.27 | 4.03 |
| 26.24 | 26.29 | 3.66 |
| 24.82 | 25.66 | 3.35 |
| 21.76 | 24.16 | 2.68 |
| 20.29 | 23.48 | 2.36 |
| 18.98 | 22.81 | 2.13 |
| 17.85 | 22.26 | 1.95 |
| 16.66 | 21.7 | 1.77 |
| 15.29 | 21.09 | 1.59 |
| 14.49 | 20.79 | 1.51 |
| 13.64 | 20.51 | 1.38 |
| 12.29 | 20.15 | 1.22 |
| 11.38 | 19.87 | 1.09 |

## 11. Correlation between $\rho_{s0}$ and $A_1^{\square}$ in other superconducting systems

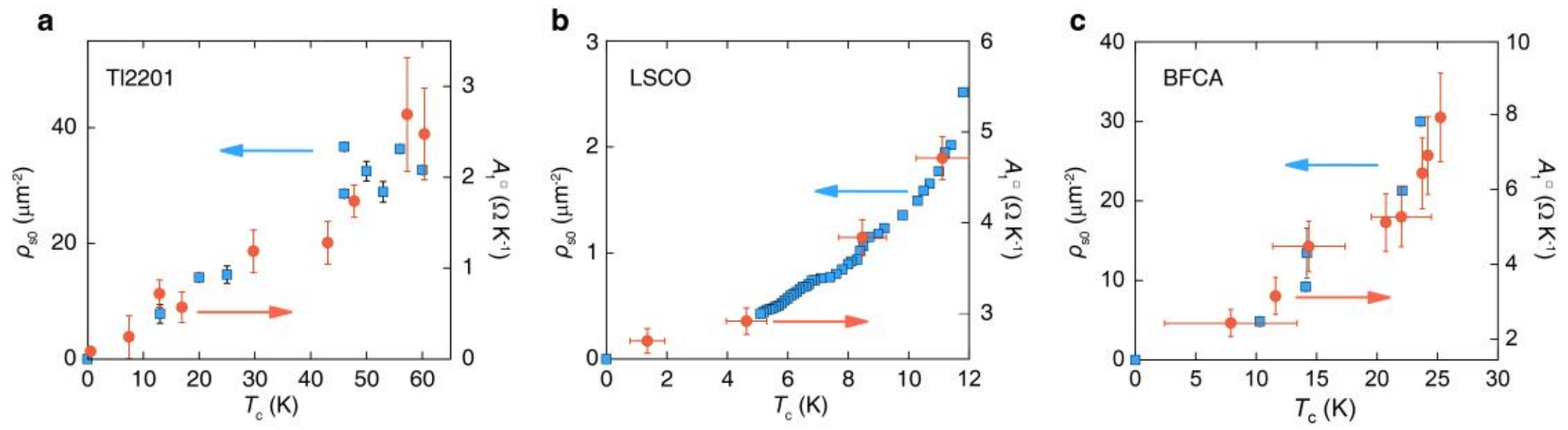

**Fig. S10. Doping dependences of $\rho_{s0}$ and $A_1^{\square}$ in a, $Tl_2Ba_2CuO_{6+\delta}$ (Tl2201), b, $La_{2-x}Sr_xCuO_4$ (LSCO), and c, $Ba(Fe_{1-x}Co_x)_2As_2$ (BFCA).** The $T_c$ values correspond to different hole dopings for **a,** and **b,** and different cobalt concentrations for **c**. Data for Tl2201 are extracted from Ref. 17 and references therein; Data for LSCO are extracted from Refs. 9,16; Data for BFCA are extracted from Refs. 18,19.

## 12. $\rho_{s0}$ versus $A_1^{\square}$ relation in overdoped LSCO

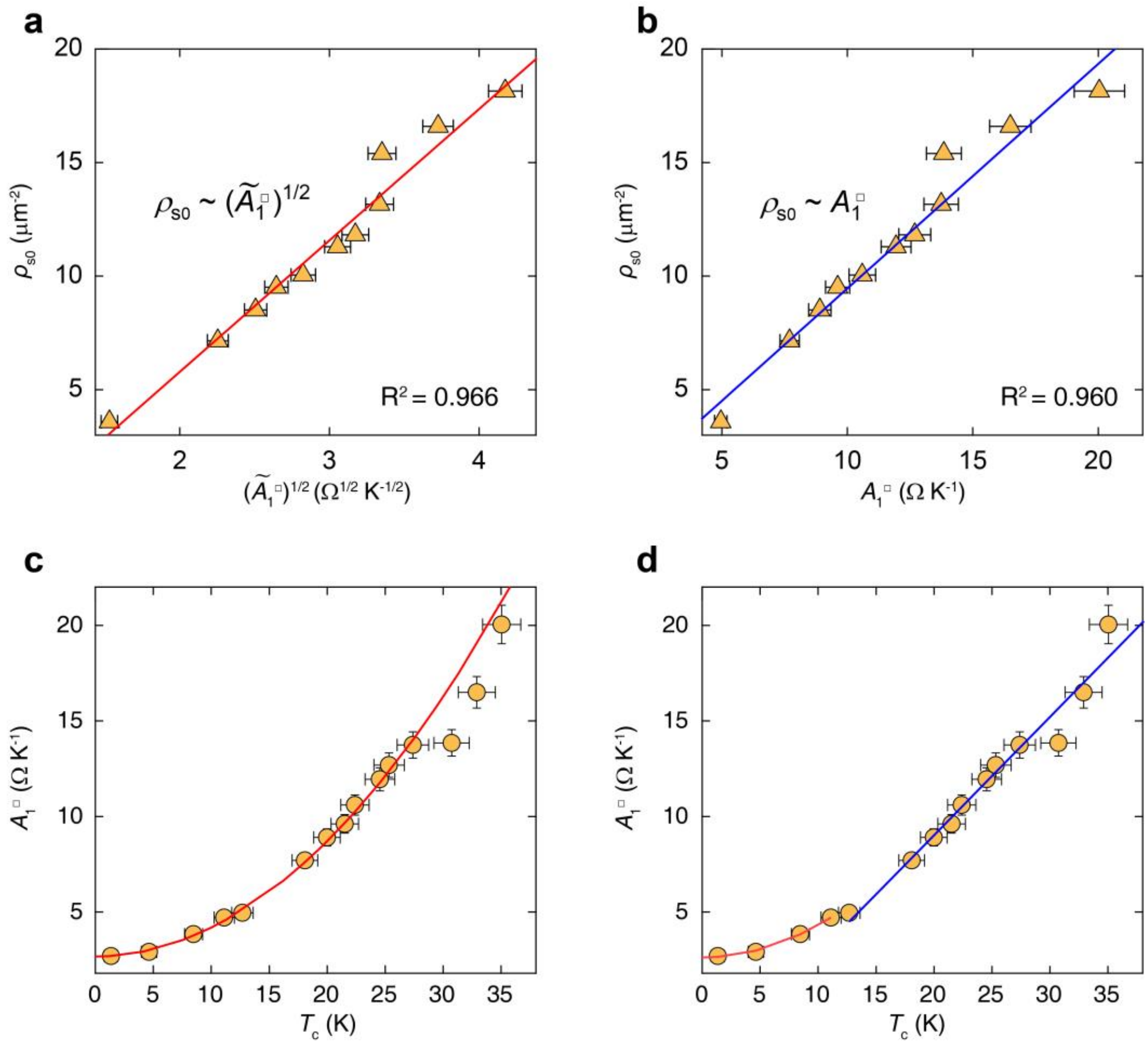

**Fig. S11. Fitting the $\rho_{s0}$ and $A_1^{\square}$ relation for overdoped LSCO with $T_c$ > 12 K using a, $\rho_{s0} \sim (\widetilde{A}_1^{\square})^{1/2}$ and b, $\rho_{s0} \sim \widetilde{A}_1^{\square}$.** It is noteworthy that the R-square ($R^2$) for two fits are almost equal, i.e., $R^2$ = 0.966 and 0.960, respectively, indicating that a quantitative relation cannot be pinned down. This is due to the large uncertainty of the $A_1^{\square}$ data for $T_c$ > 12 K, where the fit of **c,** $T_c \sim (\widetilde{A}_1^{\square})^{1/2}$ or **d,** $T_c \sim A_1^{\square}$ [that is, similar to the behavior of $T_c(\rho_{s0})$ which changes from parabolic to linear relation as $T_c$ exceed 12 K (Ref. 9)] are better remains unclear. Further studies are necessary to quantify the relation in this range. Data are obtained from Refs. 9,16.

## 13. Residual resistivity ratio of ion-gated FeSe and overdoped Tl2201

Figure S12 shows the evolution of $RRR = \rho(200\ \mathrm{K})/\rho_0$ for ion-gated FeSe and overdoped Tl2201, both of which exhibit positive correlation with $T_c$, indicating that doping changes the amount of disorder in these systems. For FeSe, the disorder may relate to the Fe vacancy, of which the effects on transport and superconducting properties were found to be suppressed by electron doping[20,21]. For overdoped cuprates, it has been believed that the dopant ion serves as an intrinsic disorder (see e.g., Ref. 22).

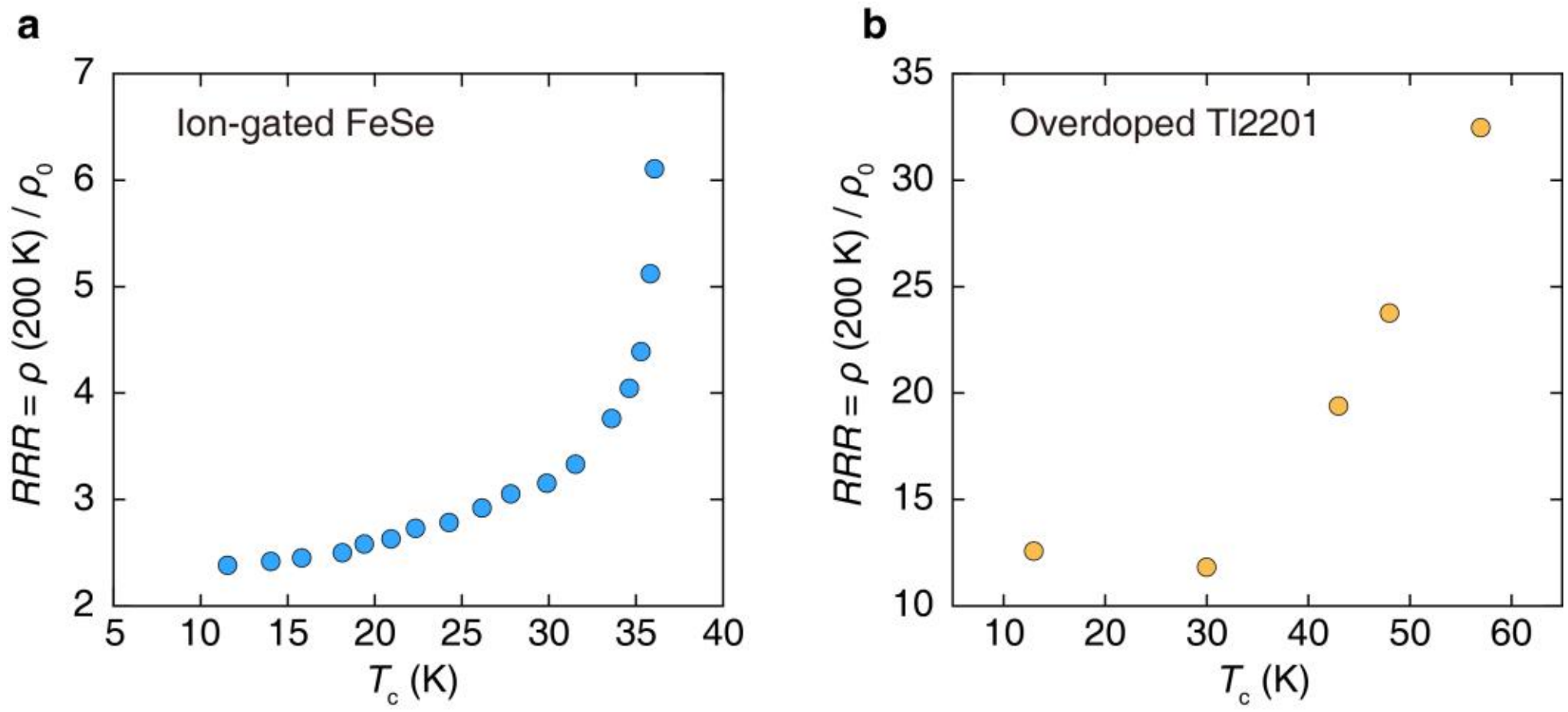

**Fig. S12. Residual resistivity ratio of $\boldsymbol{RRR = \rho(200\ \mathrm{K})/\rho_0}$, where $\boldsymbol{\rho_0}$ is obtained by fitting the resistivity above superconducting transition with $\boldsymbol{\rho = \rho_0 + A_1 T}$. a,** Ion-gated FeSe film. **b,** Overdoped $Tl_2Ba_2CuO_{6+\delta}$ (Tl2201). The $T_c$ values correspond to different electron dopings for **a,** and different hole dopings for **b,**. Data for overdoped Tl2201 are obtained from Ref. 23 and references therein

## 14. Two-dimensional Yukawa-Sachdev-Ye-Kitaev model

This supplementary section will consider the 2d-YSYK model[24-26] with the potential disorder $v$ and the spatially random Yukawa coupling $g'$. We start with the imaginary time action

$$S = \int d\tau \sum_{\boldsymbol{k},\sigma=\pm1} \sum_{i=1}^{N} \psi_{i\boldsymbol{k}\sigma}^{\dagger}(\tau)\,[\partial_\tau + \epsilon(\boldsymbol{k})]\psi_{i\boldsymbol{k}\sigma}(\tau) + \frac{1}{2}\int d\tau \sum_{\boldsymbol{q}} \sum_{i=1}^{N} \phi_{i\boldsymbol{q}}(\tau)\,[-\partial_\tau^2 + \omega^2(\boldsymbol{q})]\phi_{i,-\boldsymbol{q}}(\tau)$$

$$+ \int d\tau d^2 r \sum_{i,j=1}^{N} \sum_{\sigma=\pm1} \frac{v_{ij}(\boldsymbol{r})}{\sqrt{N}} \psi_{i\sigma}^{\dagger}(\boldsymbol{r},\tau)\psi_{j\sigma}(\boldsymbol{r},\tau) + \int d\tau d^2 r \sum_{i,j,l=1}^{N} \sum_{\sigma=\pm1} \frac{g'_{ijl}(\boldsymbol{r})}{N} \psi_i^{\dagger}(\boldsymbol{r},\tau)\psi_j(\boldsymbol{r},\tau)\phi_l(\boldsymbol{r},\tau), \tag{S3}$$

where the potential disorder $v_{ij}$ and interaction disorder $g'_{ijl}$ satisfy

$$\overline{v_{ij}(\boldsymbol{r})} = 0, \qquad \overline{v_{ij}^{*}(\boldsymbol{r})v_{i'j'}(\boldsymbol{r}')} = v^2\delta(\boldsymbol{r}-\boldsymbol{r}')\delta_{ii'}\delta_{jj'},$$

$$\overline{g'_{ijl}(\boldsymbol{r})} = 0, \qquad \overline{g'^{*}_{ijl}(\boldsymbol{r})g'_{i'j'l'}(\boldsymbol{r}')} = g'^2\delta(\boldsymbol{r}-\boldsymbol{r}')\delta_{ii'}\delta_{jj'}\delta_{ll'}. \tag{S4}$$

This was the model studied in Ref. 21, with an emphasis on the $v = 0$ case. The lattice dispersion is chosen as

$$\epsilon(\boldsymbol{k}) = -2t(\cos k_z + \cos k_y) - \mu,$$
$$\omega^2(\boldsymbol{q}) = m_{\mathrm{b}}^2 + 2J(2 - \cos q_x - \cos q_y), \tag{S5}$$

where $t$ is the fermion hopping, $\mu$ is the chemical potential, $m_{\mathrm{b}}$ is the bare boson mass and $J$ determines the boson dispersion.

A fixed length constraint

$$\sum_{\boldsymbol{q}} \sum_{i=1}^{N} \phi_{i\boldsymbol{q}}(\tau)\phi_{i,-\boldsymbol{q}}(\tau) = \frac{N}{\gamma}, \tag{S6}$$

is included and the boson mass $m_{\mathrm{b}}$ is self-consistently determined for a fixed value of $\gamma$. We tune the scalar mass $m_{\mathrm{b}}$ away from the quantum critical point, and present our results in terms of the zero-temperature value of the renormalized boson mass $M^2 = m_b^2 - \Pi(\omega = 0,\ T = 0)$, where $\Pi$ is the bosonic self-energy. The conductivity calculations[26,27] yields the residual resistivity $\rho_0 \sim v^2$ and the linear-in-$T$ slope $A_1 \sim g'^2$. Considering the fact that doping changes the amount of disorder (Fig. S6), we have also tuned the potential disorder $v$ simultaneously. For all the results presented below, we fix the following parameters: $\sqrt{J}/t = 1$, $\mu/t = -0.5$, $g'/t^{3/2} = 5$. $\gamma$ and $v$ are chosen as

$$(1/\gamma)/t = 0.19 - 0.00375(i - 1),$$
$$v/t = 0.5 + 0.075(i - 1), \tag{S7}$$

where $i$ goes from 1 to 14.

Following the steps in Ref. 25, we obtain normal state resistivity $\rho$, the superconducting transition temperature $T_{\mathrm{c}}$, and superfluid stiffness $\rho_{\mathrm{s}}$. As shown in Fig. S13b, the dependence of $T_{\mathrm{c}}$ on $\rho_{\mathrm{s0}}$ exhibits a power-law behavior, $T_{\mathrm{c}} \propto \rho_{\mathrm{s0}}^{0.67 \pm 0.03}$, which shares similarities with the measurement $T_{\mathrm{c}} \propto \rho_{\mathrm{s0}}^{0.55 \pm 0.11}$ of ion-gated FeSe (Fig. 2c). The coefficient $A_1$ is extracted by fitting the $T$-linear resistivity in Fig. S13c with $\rho = \rho_0 + A_1 T$. We find superfluid stiffness $\rho_{\mathrm{s0}}$ can increase approximately linearly with the slope $A_1$, as shown in Fig. 4b in the main text.

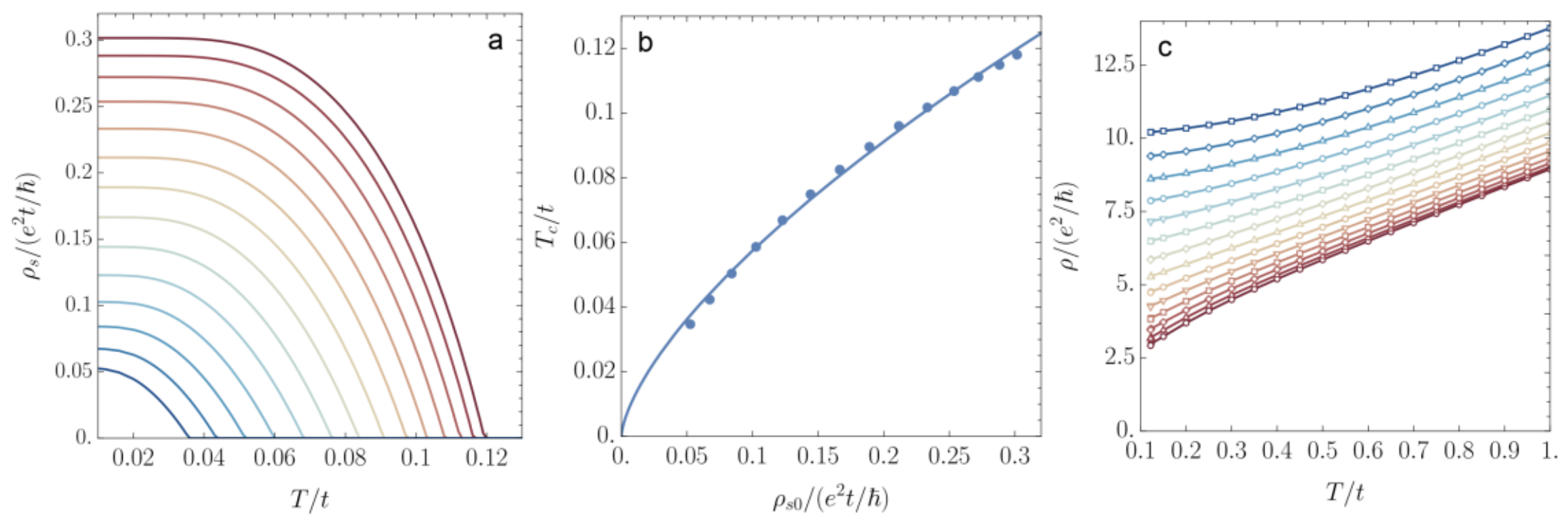

**Fig. S13. Resistivity and superfluid stiffness in 2d-YSYK model. a,** Temperature dependence of superfluid stiffness for different values of $M/t$. **b,** Dependence of $T_c$ on the zero-temperature superfluid density, $\rho_{s0}$ extracted from a. The solid line is the fit to $T_c = \gamma \rho_{s0}^{0.67}$. **c,** Normal state resistivity for varies $M/t$. In a and **c,** from dark red ($i$ = 1) to dark blue ($i$ = 14) the system is tuned away from the critical point, for $M/t$ = 0.4, 0.61, 0.79, 0.95, 1.1, 1.25, 1.39, 1.53, 1.67, 1.81, 1.95, 2.09, 2.23, 2.37.

## 15. The residual *T*-linear resistivity term of overdoped cuprates

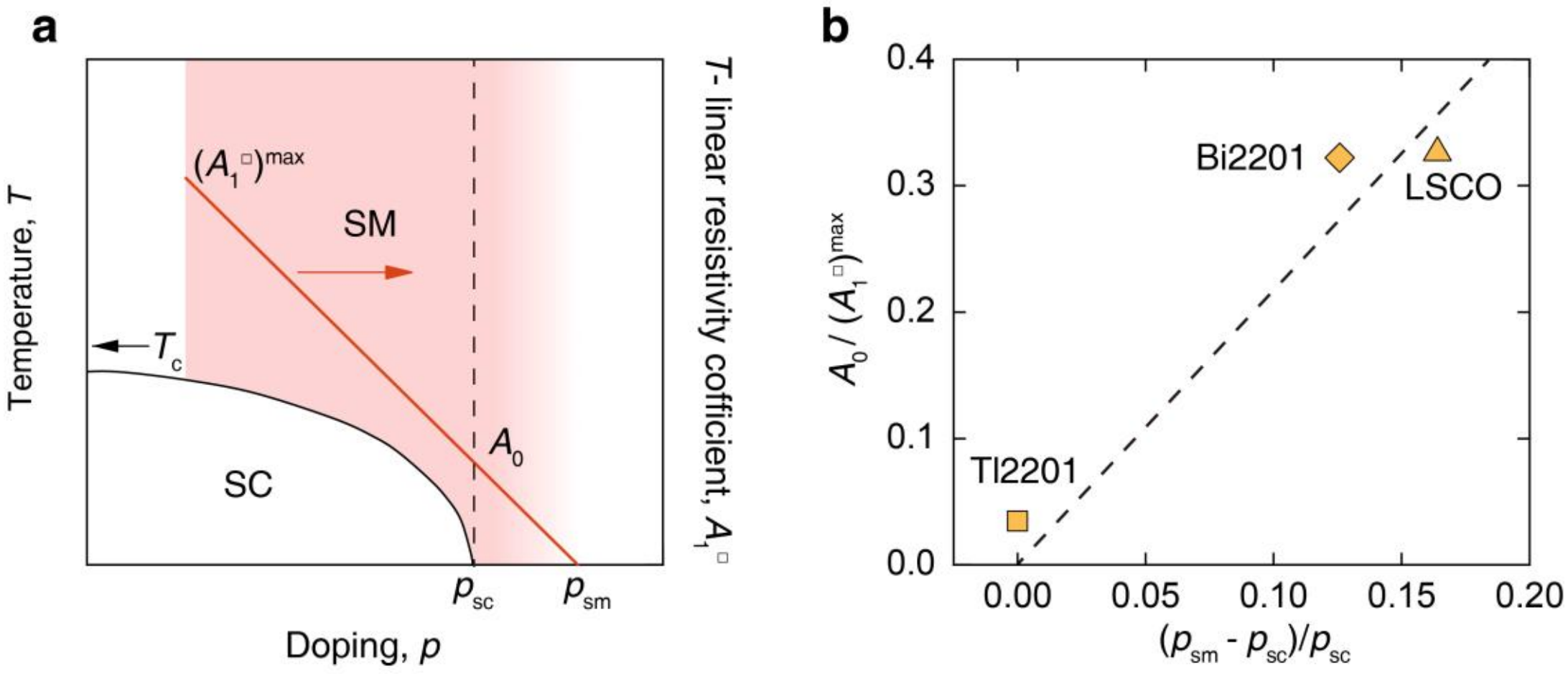

**Fig. S14. Origin of the residual *T*-linear resistivity term for hole-doped cuprates. a,** Schematic temperature versus doping phase diagram of overdoped cuprates. $T_c$ is the onset temperature of the superconducting (SC) phase, $A_1^{\square}$ is the $T$-linear resistivity coefficient of the strange-metal (SM) state. $p_{sc}$ and $p_{sm}$ represent the doping levels at which $T_c$ and $A_1^{\square}$ fade away, respectively. The mismatch between $p_{sc}$ and $p_{sm}$ results in a residual $A_1^{\square}$ value ($A_0$) at the boundary of the SC dome. **b,** Dependence of $A_0$ on the mismatch between $p_{sc}$ and $p_{sm}$ for $Tl_2Ba_2CuO_{6+\delta}$ (Tl2201), $Bi_2Sr_2CuO_{6+\delta}$ (Bi2201), and $La_{2-x}Sr_xCuO_4$ (LSCO). Data are obtained from Ref. 23.